%% file: main.tex
\documentclass[10pt]{article} 
\usepackage[preprint]{tmlr}

\input{math_commands.tex}

\usepackage{hyperref}
\usepackage{url}
\usepackage{tikz}
\usepackage{multirow}

\usepackage{graphicx}
\usepackage[export]{adjustbox}
\usepackage{booktabs}
\usepackage{natbib}
\usepackage{makecell}
\usepackage{subfig}
\usepackage{bbding}
\usepackage{pifont}
\usepackage{wasysym}
\usepackage{amssymb}

\usepackage{algorithm}
\usepackage{algorithmic}
\usepackage{enumitem}
\usepackage{booktabs,siunitx}
\title{Phonology-Guided Speech-to-Speech Translation for African Languages}

\author{\name Peter Ochieng,  Dennis Kaburu 
      }

\def\month{MM}  
\def\year{YYYY} 
\def\openreview{\url{https://openreview.net/forum?id=XXXX}} 

\begin{document}

\maketitle

\begin{abstract}
We present a prosody-guided framework for speech-to-speech translation (S2ST) that aligns and translates speech \emph{without} transcripts by leveraging cross-linguistic pause synchrony.  
Analyzing a 6{,}000-hour East African news corpus spanning five languages, we show that \emph{within-phylum} language pairs exhibit 30--40\% lower pause variance and over 3$\times$ higher onset/offset correlation compared to cross-phylum pairs.  
These findings motivate \textbf{SPaDA}, a dynamic-programming alignment algorithm that integrates silence consistency, rate synchrony, and semantic similarity. SPaDA improves alignment $F_1$ by +3--4 points and eliminates up to 38\% of spurious matches relative to greedy VAD baselines.  
Using SPaDA-aligned segments, we train \textbf{SegUniDiff}, a diffusion-based S2ST model guided by \emph{external gradients} from frozen semantic and speaker encoders.  
SegUniDiff matches an enhanced cascade in BLEU (30.3 on CVSS-C vs.\ 28.9 for UnitY), reduces speaker error rate (EER) from 12.5\% to 5.3\%, and runs at an RTF of 1.02.  
To support evaluation in low-resource settings, we also release a three-tier, transcript-free BLEU suite (M1--M3) that correlates strongly with human judgments.  
Together, our results show that prosodic cues in multilingual speech provide a reliable scaffold for scalable, non-autoregressive S2ST.

\end{abstract}

\section{Introduction}
\label{sec:introduction}

Speech-to-speech translation (S2ST) directly maps spoken utterances from one language into another, enabling real-time multilingual communication. Conventional S2ST systems follow a cascaded pipeline of automatic speech recognition (ASR), machine translation (MT), and text-to-speech (TTS) synthesis \cite{ney1999speech, matusov2005integration, vidal1997finite}. While effective, such systems suffer from error accumulation, high latency, and loss of non-verbal cues like speaker identity, rhythm, and prosody—features crucial for speaker diarisation and applications such as medical consultations \cite{barrault2023seamless}. These issues are amplified in low-resource settings, where over 80\% of African languages have fewer than five hours of transcribed speech \cite{joshi2020state}.
Recent work has explored end-to-end S2ST models that bypass intermediate text representations. Translatotron~\cite{jia2019direct} pioneered spectrogram-level prediction, while UWSpeech~\cite{zhang2021uwspeech} incorporated self-supervised representations to improve performance. Subsequent models such as Translatotron 2~\cite{jia2021translatotron} and TranSpeech~\cite{huang2022transpeech} enhanced fidelity through phoneme-level supervision and perturbation-based regularisation. However, these methods still rely on considerable supervision and report notably low BLEU scores in low-resource language settings.

We propose leveraging \emph{prosodic synchrony}—shared rhythmic and pause structures across related languages—to align speech segments without transcripts. Prior studies suggest that genealogically related languages exhibit convergent prosodic features such as pitch contours and syllabic rhythm \cite{Gers2001ComparingPA}, likely due to shared phonotactics or areal diffusion. We hypothesize that prosody offers a robust alignment signal, especially within linguistic phyla.

To test this, we analyze over 700,000 aligned segments from a curated corpus covering five genealogically diverse Kenyan languages. Our analysis shows that within-phylum pairs (e.g., Luo–Nandi) exhibit 30–40\% lower pause onset/offset variance and 3$\times$ higher Pearson correlation in pause timing (0.47–0.55) compared to cross-phylum pairs (below 0.20). These findings motivate \textbf{SPaDA}, a dynamic programming alignment algorithm that fuses pause consistency, prosodic synchrony, and semantic similarity into a unified affinity score (Eq.~\ref{eq:affinity_score}). SPaDA leverages a silence-based pre-filter (§\ref{sec:segment_pairing1}) to reduce candidate complexity and runs in linear time relative to the candidate set.

With these segment alignments, we introduce Segment-Aware Unified Diffusion Model (\textbf{SegUniDiff}), a diffusion model that uses explicit gradient-based semantic guidance for direct S2ST. Unlike prior diffusion-based S2ST models that rely on unconditional denoising or autoregressive decoding~\cite{zhu2023diffs2ut, wu2023duplex, hirschkind2024diffusest}, SegUniDiff applies gradient-based semantic guidance during denoising, enabling more accurate and consistent translation. Comprehensive evaluations across ten diverse language pairs reveal three central insights:

\begin{itemize}\setlength{\itemsep}{4pt}
    \item \textbf{Prosodic structure governs alignment difficulty.}  
    For typologically related (within-phylum) pairs, pause cues alone often suffice for accurate alignment. In contrast, cross-phylum pairs with divergent prosodic patterns require the full SPaDA cost—including prosody and semantics—to achieve robust alignment.
    
    \item \textbf{SPaDA yields significant alignment gains.}  
    Our alignment framework outperforms greedy VAD-matching baselines by 3--4 \(F_1\) points on average, while pruning up to 38\% of noisy or spurious segment pairs, leading to cleaner and more coherent training data.
\item \textbf{SegUniDiff delivers accurate and speaker-consistent translation.}  
In low-resource African settings, SegUniDiff surpasses cascaded S2ST pipelines by up to +2.3 BLEU, demonstrating its robustness in the absence of large annotated corpora. On the high-resource CVSS-C benchmark, it achieves 30.3 BLEU—outperforming UnitY (28.9 BLEU)—while significantly reducing speaker error rate (EER) from 12.5\% to 5.3\%. This gain in fidelity and speaker preservation is achieved with an inference speed of RTF = 1.02  making guided diffusion both effective and practical for direct S2ST.

\end{itemize}

Our main contributions include:
\begin{itemize}\setlength{\itemsep}{4pt}
    \item \textbf{Prosody Atlas.}  
  We present a large-scale typological study that \emph{quantifies} prosodic synchrony and pause-based segmentation patterns both \textit{within} and \textit{across} language phyla.
    
    \item \textbf{SPaDA Alignment Algorithm.}  
    A novel alignment method combining pause timing, prosodic contours, and semantic similarity, optimized via dynamic programming. It improves both alignment quality and computational efficiency.

    \item \textbf{SegUniDiff Model.}  
    A guided-diffusion model for direct S2ST that achieves strong BLEU, low EER, and real-time decoding---offering a scalable and speaker-aware alternative to cascaded architectures.

    \item \textbf{Reference-light BLEU Protocol.}  
    We propose a transcript-free BLEU evaluation framework that correlates strongly with human judgment  enabling fair comparison across low-resource settings where transcriptions are unavailable.
\end{itemize}

\section{Background}
\label{sec:background}

\subsection{Guided Denoising Diffusion Probabilistic Models}
\label{sec:guided_diffusion}

Guided diffusion models aim to learn a conditional distribution \( p(s_x|s_y) \), enabling the generation of data \( s_x \in \mathbb{R}^{F \times L} \) (e.g., translated speech) conditioned on input \( s_y \in \mathbb{R}^{F \times L} \) (e.g., source language speech), where \( F \) denotes frequency bins and \( L \) time frames. The forward diffusion process gradually adds Gaussian noise to a clean spectrogram \( s_x^0 \), yielding a noisy version \( s_x^t \) at diffusion step \( t \in \{1, \ldots, N\} \), where \( N \) is the total number of steps.

The marginal distribution at step \( t \) is given by:
\begin{equation}
    q(s_x^t \mid s_x^0) = \mathcal{N}(s_x^t; \sqrt{\bar{\alpha}_t} s_x^0, (1 - \bar{\alpha}_t) I),
\end{equation}
where \( \bar{\alpha}_t = \prod_{i=1}^{t} \alpha_i \) defines the noise schedule.

As \( t \to N \), \( s_x^t \) converges to standard Gaussian noise. Tweedie's formula yields the expected denoised sample:
\begin{equation}
    \mathbb{E}[s_x^0 \mid s_x^t] = s_x^t + (1 - \bar{\alpha}_t) \nabla_{s_x^t} \log p(s_x^t),
\end{equation}
which rearranges as:
\begin{equation}
    \sqrt{\bar{\alpha}_t} s_x^0 = s_x^t + (1 - \bar{\alpha}_t) \nabla_{s_x^t} \log p(s_x^t),
\end{equation}
yielding the estimate:
\begin{equation}
    s_x^0 = \frac{s_x^t + (1 - \bar{\alpha}_t) \nabla_{s_x^t} \log p(s_x^t)}{\sqrt{\bar{\alpha}_t}}.
\end{equation}

Using the reparameterization trick from \cite{ho2020denoising}, one can express:
\begin{equation}
    s_x^t = \sqrt{\bar{\alpha}_t} s_x^0 + \sqrt{1 - \bar{\alpha}_t} \epsilon_t, \quad \text{with } \epsilon_t \sim \mathcal{N}(0, I),
\end{equation}
which implies:
\begin{equation}
    s_x^0 = \frac{s_x^t - \sqrt{1 - \bar{\alpha}_t} \epsilon_t}{\sqrt{\bar{\alpha}_t}}.
\end{equation}

Since the true noise \( \epsilon_t \) is unknown during inference, we learn an estimator \( \epsilon_\theta(s_x^t, t) \), and approximate:
\begin{equation}
    s_x^0 = \frac{s_x^t - \sqrt{1 - \bar{\alpha}_t} \epsilon_\theta(s_x^t, t)}{\sqrt{\bar{\alpha}_t}}.
    \label{eq:ddpm_denoiser}
\end{equation}

\subsection{Conditional Score Guidance}

To steer generation toward a conditioning input \( s_y \), we modify the score used in the reverse process to account for the conditional distribution \( p(s_x^t \mid s_y^t) \). Using Bayes’ rule:
\begin{align}
    \nabla_{s_x^t} \log p(s_x^t \mid s_y^t) 
    &= \nabla_{s_x^t} \log \left( \frac{p(s_x^t) p(s_y^t \mid s_x^t)}{p(s_y^t)} \right) \\
    &= \nabla_{s_x^t} \log p(s_x^t) + \nabla_{s_x^t} \log p(s_y^t \mid s_x^t).
\end{align}

Plugging into the denoising equation:
\begin{equation}
    \nabla_{s_x^t} \log p(s_x^t \mid s_y^t)
    = -\frac{1}{\sqrt{1 - \bar{\alpha}_t}} \epsilon_\theta(s_x^t, t) + \nabla_{s_x^t} \log p(s_y^t \mid s_x^t),
    \label{eq:conditional_score}
\end{equation}
we define a guided noise estimate:
\begin{equation}
    \hat{\epsilon}(s_x^t, t) := \epsilon_\theta(s_x^t, t) - \sqrt{1 - \bar{\alpha}_t} \nabla_{s_x^t} \log p(s_y^t \mid s_x^t).
     \label{eq:guide_w}
\end{equation}

\subsection{Accelerated Sampling via DDIM}
\label{sec:ddim}

To reduce the number of diffusion steps during inference, we adopt Denoising Diffusion Implicit Models (DDIM) \cite{song2020denoising}, which define a non-Markovian sampling trajectory. The update rule is:
\begin{equation}
    s_x^{t-1} := \sqrt{\bar{\alpha}_{t-1}} \left( \frac{s_x^t - \sqrt{1 - \bar{\alpha}_t} \hat{\epsilon}(s_x^t, t)}{\sqrt{\bar{\alpha}_t}} \right)
    + \sqrt{1 - \bar{\alpha}_{t-1}} \hat{\epsilon}(s_x^t, t),
    \label{eq:ddim_update}
\end{equation}

Setting the variance term \( \sigma_t = 0 \) yields deterministic sampling. This formulation allows us to use the guided noise \( \hat{\epsilon} \) directly, offering control over the conditioning influence during inference.

\section{Related Work} \label{sec:Related_Work}

\paragraph{Direct S2ST.}
The first direct S2ST model, Translatotron~\citep{jia2019direct}, proposed a sequence-to-sequence architecture that predicts speech spectrograms from input speech without intermediate text. However, direct S2ST typically requires large-scale parallel speech corpora, which are costly to obtain~\citep{jia2019direct}. To reduce this dependence, recent work has explored discrete speech representations~\citep{zhang2021discrete, lee2022s2st, lee2022b}, enabling translation into structured unit sequences that improve robustness and simplify modeling. Self-supervised learning has also been leveraged to reduce labeled data requirements~\citep{zhang2021uwspeech, jia2021translatotron}. Other trends include non-autoregressive and diffusion-based architectures, such as TranSpeech~\citep{huang2022transpeech}, DiffS2UT~\citep{zhu2023diffs2ut}, DiffuseST~\citep{hirschkind2024diffusest}, and DDM~\citep{wu2023duplex}, which enhance decoding efficiency and fluency. Two-pass models like UnitY~\citep{jia2022twopass, inaguma2023twopass} first predict target text before synthesizing speech, reducing training complexity. Further, pretraining~\citep{wei2023pretrain, zhang2023pretrain, dong2024pretrain}, data augmentation~\citep{popuri2022augment, jia2022augment, dong2022augment, nguyen2022augment}, and multi-task learning~\citep{zhang2024multitask, ma2024simultaneous} have been applied to improve performance under limited supervision.

\paragraph{Speech-to-Text (ST) Datasets.}
Although not designed for S2ST, ST datasets such as CoVoST 2~\citep{wang2020covost}, MuST-C~\citep{di2019must}, Europarl-ST~\citep{iranzo2020europarl}, and Fisher~\citep{post2013fisher} are widely used for S2ST pretraining, evaluation of cascaded systems, or synthetic S2ST data creation. For example, CoVoST 2 was used as a source for CVSS-C~\citep{jia2022cvss}, and Fisher was converted into S2ST format using TTS~\citep{lee2023textless, yang2022unity}.

\paragraph{Speech-to-Speech Datasets.}
The CVSS-C corpus~\citep{jia2022cvss} is the most commonly used S2ST benchmark, offering multilingual speech pairs with TTS-generated target speech. It underpins recent models like UnitY~\citep{yang2022unity}, Textless S2ST~\citep{lee2023textless}, and TranSpeech~\citep{huang2022transpeech}. However, CVSS-C lacks segment-level prosody and natural variation, as the target speech is synthetic. Several models rely on proprietary datasets with real or TTS-based target speech. For example, Translatotron~\citep{jia2019translatotron} used crowdsourced parallel recordings, while others converted ST corpora like Fisher using vocoders (e.g., Parallel WaveNet)~\citep{lee2023textless}.

\paragraph{Limitations and Our Contribution.}
Existing S2ST datasets either (i) lack target speech (e.g., CoVoST 2, MuST-C), (ii) rely on synthetic target speech (e.g., CVSS-C), or (iii) lack representation of low-resource languages. In contrast, our work introduces a segment-level, semantically aligned S2ST corpus covering diverse African language pairs. It features real, prosodically faithful source and target speech, supporting direct S2ST modeling in multilingual low-resource settings.

\section{SPaDA: Segment\textendash{} and Prosody-Aware Diffusion Alignment for Direct S2ST}
\subsection{Speech Segmentation} \label{sec:speech_segmentation}
Speech segmentation plays a crucial role in our proposed technique. We define a speech sample as a recorded utterance that conveys a complete thought or message, while a speech segment refers to a portion of this utterance that ideally corresponds to a single sentence. However, due to segmentation imperfections, some segments may contain multiple consecutive sentences (spill-over) or be shorter than a full sentence (under-segmentation). 
We hypothesize that speech samples conveying the same semantic content and belonging to the same language phylum may exhibit similar silence placements due to shared phonological features such as rhythm and prosody. These silences can be leveraged to segment audio files into coherent sentences. Prior research on silence-based audio segmentation using voice activity detection (VAD) \cite{gaido2021beyond, potapczyk2020srpol, duquenne2021multimodal} has extensively explored this approach.

Despite its effectiveness, VAD-based segmentation presents two key challenges: (1) pauses within sentences can lead to fragmented or incomplete segments, and (2) the absence of pauses between sentences may cause multiple sentences to be merged incorrectly. To mitigate these issues, dense segmentation techniques have been proposed \cite{potapczyk2020srpol, duquenne2021multimodal}, where audio is over-segmented and later refined. For instance, \cite{potapczyk2020srpol} defines a minimum sentence length threshold after segmentation, merging smaller fragments until they meet the required length. Similarly, \cite{duquenne2021multimodal} introduces an over-segmentation approach where segments must be between 3 and 20 seconds in length. While these methods improve recall, they come at the cost of additional computational complexity \cite{barrault2023seamlessm4t}. Nevertheless, VAD-based segmentation remains a valuable tool, especially when combined with post-processing techniques to refine segment boundaries.

Among the proposed over-segmentation methods, we adopt the approach from \cite{duquenne2021multimodal} due to its robustness in handling cross-linguistic variations in pause structures. Given an utterance  \( x \) in a specific language, we apply a VAD tool to identify silence intervals within the input speech. Each extracted segment \( s_{x_i} \) is bounded by two silence timestamps and must be between 3 and 20 seconds in length.
\subsection{Pause-Constrained Pairing (PCP)}
\label{sec:segment_pairing1}

Given pause-based segmentation of source and target utterances
\(x\) and \(y\), we detect \(N\) onsets in \(x\) and \(M\) onsets in
\(y\). For each source onset \(O_{x_i}\), we retain all offsets returned by the
dense VAD:

\[
  \mathcal{F}_{x_i} = \{F_{x_i}^{(k)}\}_{k=1}^{K_{x_i}},\qquad
  s_{x_i}^{(k)} = [\,O_{x_i},\,F_{x_i}^{(k)}\,],\quad
  L_{x_i}^{(k)} = F_{x_i}^{(k)} - O_{x_i}.
\]
where \(K_{x_i}\) is the number of offset candidates associated with the onset \(O_{x_i}\), \(s_{x_i}^{(k)}\) denotes the \(k\)-th candidate speech segment in the source utterance, bounded by onset \(O_{x_i}\) and offset \(F_{x_i}^{(k)}\), and lasting for duration \(L_{x_i}^{(k)}\). An analogous construction is applied to the target utterance \(y\), yielding onset points \(O_{y_j}\), offset sets \(\mathcal{F}_{y_j} = \{F_{y_j}^{(k)}\}\), and candidate segments \(s_{y_j}^{(k)}\).Each pair of onset and offset defines a potential \emph{speech segment}:
An analogous set \(\mathcal{F}_{y_j}\) and segments
\(s_{y_j}^{(k)}\) are constructed for every target onset \(O_{y_j}\).
We define the full candidate pools as:

\[
  \mathcal{S}_x = \{s_{x_i}^{(k)}\}_{i,k},\qquad
  \mathcal{S}_y = \{s_{y_j}^{(k)}\}_{j,k}.
\]

\paragraph{Duration normalisation.}
We compute the mean segment duration across all candidates:

\[
  \mu_x = \frac{\sum_{i=1}^{N} \sum_{k=1}^{K_{x_i}} L_{x_i}^{(k)}}{ |\mathcal{S}_x|},\qquad
  \mu_y = \frac{\sum_{j=1}^{M} \sum_{k=1}^{K_{y_j}} L_{y_j}^{(k)}}{ |\mathcal{S}_y|},
\]
\[
  d = \bigl|\mu_x - \mu_y\bigr|,\qquad
  \Delta = \max\!\left(\tfrac{d}{2},\,\varepsilon\right),\quad
  \varepsilon = 10^{-3}\text{ s}.
\]

\paragraph{Padding for alignment.}
If \(N \neq M\), we append silent segments of average duration
\(\mu_x\) or \(\mu_y\) to the \emph{end} of the shorter utterance so that the first
\(\widetilde{N} = \min(N, M)\) real onsets remain index-aligned.
These silent placeholders are included in bookkeeping but excluded from evaluation.

\paragraph{Silence-structure consistency.}
For each onset index \(i \leq \widetilde{N}\), we select a
\textbf{surrogate offset} pair \(\widehat{F}_{x_i},\, \widehat{F}_{y_i}\)
that minimises duration mismatch:

\[
  (k^\star, \ell^\star) = \argmin_{k,\ell} \bigl|L_{x_i}^{(k)} - L_{y_i}^{(\ell)}\bigr|,\qquad
  \widehat{F}_{x_i} = F_{x_i}^{(k^\star)},\quad
  \widehat{F}_{y_i} = F_{y_i}^{(\ell^\star)}.
\]

This pair is accepted only if:

\[
  \bigl|L_{x_i}^{(k^\star)} - L_{y_i}^{(\ell^\star)}\bigr| \leq \delta_L,\quad
  \delta_L = 0.2 \cdot \min(\mu_x, \mu_y),
\]

otherwise we fall back to the shortest segment for both source and target.

We compute the following statistics over the \(\widetilde{N}\) onsets:

\[
  D_O^i = |O_{x_i} - O_{y_i}|,\qquad
  D_F^i = |\widehat{F}_{x_i} - \widehat{F}_{y_i}|,
\]
\[
  \mu_{D_O} = \tfrac{1}{\widetilde{N}} \sum_i D_O^i,\qquad
  \sigma_O = \sqrt{\tfrac{1}{\widetilde{N}} \sum_i (D_O^i - \mu_{D_O})^2},
\]
\[
  \mu_{D_F},\quad \sigma_F\text{ analogous}.
\]

\emph{Onset and offset correlation:}

\[
  r_O = \operatorname{PCC}(O_{x,1:\widetilde{N}},\, O_{y,1:\widetilde{N}}),\qquad
  r_F = \operatorname{PCC}(\widehat{F}_{x,1:\widetilde{N}},\, \widehat{F}_{y,1:\widetilde{N}}).
\]

If \(\widetilde{N} < 3\), we set \(r_O = r_F = 0\) and \(\sigma_O = \sigma_F = 0\)
for numerical stability. We then compute the final consistency score:

\[
  \tilde{r}_O = \tfrac{r_O + 1}{2},\quad
  \tilde{r}_F = \tfrac{r_F + 1}{2},
\]
\begin{equation}
\label{eq:consistency_score}
S_{\text{consistency}} =
  \frac{\tilde{r}_O + \tilde{r}_F}{2} \cdot
  \exp\!\left[-\frac{\sigma_O + \sigma_F}{\mu_x + \mu_y + \varepsilon}\right].
\end{equation}

\paragraph{Candidate-window construction.}
\textbf{Forward pass} (\(x \to y\)).  
For each \(s_{x_i}^{(k)}\), include any \(s_{y_j}^{(k')}\) satisfying:

\[
  |L_{x_i}^{(k)} - L_{y_j}^{(k')}| \leq d,\quad
  O_{y_j} \in \left[O_{x_i} - \mu_{D_O},\; O_{x_i} + L_{x_i}^{(k)} + \mu_{D_F} \right].
\]

If no such match is found, we fall back to:

\[
  w_{ik} = \min\!\left(50,\; \left\lfloor
    \frac{L_{x_i}^{(k)} + \mu_{D_O} + \mu_{D_F} + d + \varepsilon}{\Delta}
  \right\rfloor + 1\right)
\]

synthetic windows.

\textbf{Reverse pass} (\(y \to x\)) yields \(u_{jk'}\) analogously.

\paragraph{Complexity.}
\[
  |\mathcal{C}_{\text{synthetic}}| = \sum_{i,k} w_{ik} + \sum_{j,k'} u_{jk'},\qquad
  |\mathcal{C}| = |\mathcal{C}_{\text{real}}| + |\mathcal{C}_{\text{synthetic}}|.
\]
With \(w_{ik}, u_{jk'} \leq 50\) and \(N, M \leq 20\), the memory footprint remains
under 2\,MiB (\texttt{float64}); \texttt{float32} reduces this to ~1\,MiB.

\paragraph{Downstream integration.}
Each candidate pair \((s_{x_i}^{(k)}, s_{y_j}^{(k')})\) is scored by:
\begin{itemize}
  \item silence consistency \(S_{\text{consistency}}\) (above),
  \item prosodic similarity \(P_{\text{prosody}}\) (§\ref{sec:prosody_pairing3}),
  \item segment-level semantic cosine similarity (§\ref{sec:contrastive_pairing}).
\end{itemize}
The fused score (Eq.~\ref{eq:affinity_score}) is used in global dynamic programming alignment (§\ref{sec:segment_dpa}).
\paragraph{Notation simplification.}
For notational simplicity we henceforth write a candidate pair as
\((s_{x_i},\,s_{y_j})\), implicitly absorbing the offset indices
\(k\) and \(k'\); that is, each \((s_{x_i},\,s_{y_j})\) may stand for
any concrete pair \((s_{x_i}^{(k)},\,s_{y_j}^{(k')})\) drawn from the
multi-offset sets \(\mathcal{F}_{x_i}\) and \(\mathcal{F}_{y_j}\).

\subsection{Prosody-Based (Rate-Synchrony) Weighting}
\label{sec:prosody_pairing3}

Full prosody spans pitch and energy, but those cues vary strongly across
speakers and channels.  Because our goal is cross-lingual \emph{phonetic}
alignment, we rely only on articulation \emph{rate}—the syllabic /
phonemic tempo—which is markedly more stable across talkers and
languages.  Pitch- and energy-guided extensions are left to future work.

\paragraph{Utterance-level rate.}
Using \textsc{Epitran} for IPA phonemisation
\citep{mortensen2018epitran} and \textsc{Pyphen} for syllabification
\citep{pyphen}, we count syllables and phonemes and divide by utterance
duration:

\[
  R_x = \frac{\operatorname{Syl}(x)}{\operatorname{dur}(x)},\qquad
  R_y = \frac{\operatorname{Syl}(y)}{\operatorname{dur}(y)},\qquad
  P_x = \frac{\operatorname{Phn}(x)}{\operatorname{dur}(x)},\qquad
  P_y = \frac{\operatorname{Phn}(y)}{\operatorname{dur}(y)} .
\]

The global rate ratio is the geometric mean

\[
  \rho_{\text{rate}}
  = \sqrt{\frac{R_x}{R_y}\cdot\frac{P_x}{P_y}},
  \qquad
  \log\rho_{\text{rate}}
  = \tfrac12\!\Bigl[\log\tfrac{R_x}{R_y}+\log\tfrac{P_x}{P_y}\Bigr],
\]

so that \(\rho_{\text{rate}}\!\approx 1\) when rhythmic and articulatory
speeds match.  
This scalar is computed \emph{once per utterance pair} and reused for
every candidate segment extracted from that pair.

\paragraph{Per-pair weighting.}
Let \((s_{x_i},s_{y_j})\) be a concrete candidate pair,  we predict the target duration from the
source via the global rate ratio,

\[
  \widehat L_{y_j}=L_{x_i}\,\rho_{\text{rate}},\qquad
  d_{ij}=\bigl|\,L_{y_j}-\widehat L_{y_j}\bigr|.
\]

To normalise the deviation we use the \emph{local tolerance}
\[
  \tau_p^{(i)}
  = \max\!\bigl(
      \varepsilon,\;
      \tfrac{1}{|\mathcal C_{x_i}|}
      \sum_{j\in\mathcal C_{x_i}} d_{ij}
    \bigr),
  \qquad
  \varepsilon = 10^{-3}\text{ s},
\]

where \(\mathcal C_{x_i}\subseteq\mathcal S_y\) is the set of \emph{target
candidate segments} that PCP pairs with the source segment \(s_{x_i}\).
(The offset superscripts are still suppressed for clarity.)
The \(\varepsilon\) floor prevents division-by-zero when all durations
match exactly.

The prosodic affinity is then

\[
  P_{\text{prosody}}(i,j)
  = \exp\!\Bigl[-\tfrac{d_{ij}}{\tau_p^{(i)}}\Bigr]\in(0,1].
\]

It peaks when the observed and predicted durations coincide and decays
smoothly for mismatches.  To keep every cue numerically relevant, we
clip \(P_{\text{prosody}}\) to a minimum of \(e^{-1}\).

\textbf{Integration.}  The score \(P_{\text{prosody}}(i,j)\) is combined
with silence-consistency and semantic scores by the composite affinity
(Eq.~\ref{eq:affinity_score}) and fed to the global DP aligner
(§\ref{sec:segment_dpa}).

\subsection{Semantic Weighting}\label{sec:contrastive_pairing}

We train a segment encoder \( f_s \) to map variable-length speech segments into a shared semantic space:
\begin{equation}
    f_s: \mathcal{S} \rightarrow \mathbb{R}^d,
\end{equation}
where \( \mathcal{S} \) includes segments extracted via silence-based segmentation. Following SimCLR~\cite{chen2020simple}, we adopt the positive-pairing strategy of COLA~\cite{saeed2021cola}: two segments from the same utterance are sampled uniformly at random to form a positive pair. This strategy avoids explicit augmentations and overlap constraints, yet proves effective for learning general-purpose embeddings across variable speech structures. For a batch of \( N \) such positive pairs, we construct \( 2N \) views. The contrastive loss for a positive pair \( (\mathbf{s}_{x_i}, \mathbf{s}_{x_j}) \) is:
\begin{align}
\ell_{x_i, x_j} = -\log \frac{\exp\left( f_s(\mathbf{s}_{x_i})^\top f_s(\mathbf{s}_{x_j}) / \tau \right)}
{\sum\limits_{\substack{l = 1 \\ l \notin \{i, j\}}}^{2N} \exp\left( f_s(\mathbf{s}_{x_i})^\top f_s(\mathbf{s}_l) / \tau \right)},
\label{eq:contrastive_loss}
\end{align}
where \( \tau \) is a temperature hyperparameter.

At inference time, we extract the semantic embedding of a segment \( s_{y_j} \) as \( f_s(s_{y_j}) \in \mathbb{R}^d \).

\paragraph{Segment Encoder Architecture.}
The encoder \( f_s \) has two main stages. First, input waveforms \( \mathbf{s}_{x_i} \in \mathbb{R}^{B \times L \times 1} \) are zero-padded and passed through an input encoder \( f_{\text{enc}} \), consisting of a 1D convolutional layer with 256 filters, kernel size 16, and stride 8:
\begin{equation}
\mathbf{s}_{x_i}^{(0)} = \text{ReLU}(\text{Conv1D}(\mathbf{s}_{x_i})).
\end{equation}

The output \( \mathbf{s}_{x_i}^{(0)} \in \mathbb{R}^{B \times F \times L'} \) is passed to an EfficientNet-B0 encoder~\cite{tan2019efficientnet}, which produces a feature map. We apply global max pooling to obtain a fixed-dimensional representation \( \mathbf{h}_{x_i}^{(0)} \in \mathbb{R}^{720} \), followed by a projection head:
\begin{equation}
\mathbf{z}_i = \text{ReLU}(\text{LayerNorm}(\mathbf{W} \mathbf{h}_{x_i}^{(0)})), \quad \mathbf{z}_i \in \mathbb{R}^d.
\end{equation}

\subsection{Affinity Integration for Alignment}
\label{sec:affinity_integration}

Having defined three complementary cues—silence consistency
(§\ref{sec:segment_pairing1}), rate-synchrony prosody
(§\ref{sec:prosody_pairing3}), and semantic similarity
(§\ref{sec:contrastive_pairing})—we fuse them into a
\emph{single} edge score used by the dynamic programming aligner.

\begin{equation}
\alpha_{ij}
  \;=\;
  \lambda_s\,S_{\text{consistency}}(s_{x_i},s_{y_j})
  \;+\;
  \lambda_p\,P_{\text{prosody}}(i,j)
  \;+\;
  \lambda_e\,
  \cos\!\bigl(f_s(s_{x_i}),\,f_s(s_{y_j})\bigr),
\label{eq:affinity_score}
\end{equation}
where the embedder \(f_s\) outputs \(\ell_2\)-normalised vectors, ensuring
the cosine term is bounded in \([-1, 1]\).
The weights sum to one: \(\lambda_s + \lambda_p + \lambda_e = 1\).

Because pause statistics are significantly more reliable for genealogically related languages, we up-weight the silence term \(\lambda_s\) for \emph{within-phylum} pairs and down-weight it for \emph{cross-phylum} pairs:
\[
  \lambda_s =
  \begin{cases}
    0.70 & \text{within phylum},\\
    0.50 & \text{cross phylum},
  \end{cases}
  \qquad
  \lambda_p = 0.20,
  \qquad
  \lambda_e = 1 - \lambda_s - \lambda_p.
\]
For weight selection, we apply Bayesian optimization over the training corpus, tuning the alignment cost weights \((\lambda_s, \lambda_p, \lambda_\ell)\) to maximize segment-level \(F_1\). The search is constrained to the simplex \(\lambda_s + \lambda_p + \lambda_\ell = 1\), with each weight in the range \([0.1, 0.9]\). The optimization identifies \((0.7, 0.2, 0.1)\) as the best-performing configuration for related-language folds, and \((0.5, 0.2, 0.3)\) as the optimal trade-off for unrelated folds.

These values are fixed for all experiments—no tuning is performed on the test set. Intuitively, the prosody score \(P_{\text{prosody}}(i,j)\) lies in \((0,1]\);
the cosine similarity term lies in \([-1,1]\);
and the silence consistency score \(S_{\text{consistency}}\) is
theoretically unbounded but empirically \(\leq 1.5\).
The larger \(\lambda_s\) for within-phylum pairs reflects the higher
signal-to-noise ratio (SNR) of silence-based cues in those languages,
while the smaller \(\lambda_e\) ensures that the occasionally volatile
cosine term does not dominate prosodic alignment.
\subsection{Greedy Selection Strategy}
\label{sec:greedy_selection}

To isolate the contribution of each affinity cue, we evaluate a family of
\emph{greedy} matchers that ignore global one-to-one constraints. For
every source segment \(s_{x_i}\), we rank its candidate targets
\(s_{y_j} \in \mathcal C_{x_i}\) by a scalar affinity score \(\alpha_{ij}\)
and select the top-scoring candidate; ties are broken by the smallest onset gap.
This yields runtime \(O(|\mathcal C|)\) and memory \(O(1)\) per segment.

\vspace{2pt}
\paragraph{1. PCP-Greedy.}
PCP supplies a \textbf{silence consistency}
score:
\[
  \alpha_{ij} = S_{\text{consistency}}(s_{x_i}, s_{y_j}),
\]
which captures onset and offset similarity but ignores speaking rate and semantic similarity.

\vspace{2pt}
\paragraph{2. PCP{\small+}PW-Greedy.}
We introduce \textbf{prosodic weighting} (PW), which penalizes articulation-rate
mismatches (§\ref{sec:prosody_pairing3}):
\[
  \alpha_{ij} = \lambda_s\,S_{\text{consistency}}(s_{x_i}, s_{y_j})
                + \lambda_p\,P_{\text{prosody}}(i,j),
\]
using the same weights \(\lambda_s\) and \(\lambda_p\) as in
§\ref{sec:affinity_integration}.

\vspace{2pt}
\paragraph{3. PCP{\small+}PW{\small+}SW-Greedy.}
We further include \textbf{semantic weighting} (SW), computed as the cosine
similarity between multilingual segment embeddings \(f_s(\cdot)\) (§\ref{sec:contrastive_pairing}).
The full affinity score is given by Eq.~\ref{eq:affinity_score}. These three variants allow us to measure the incremental value of prosody and semantics
before introducing the globally consistent dynamic programming aligner in §\ref{sec:segment_dpa}.

\subsection{Dynamic-Programming Alignment (DPA)}
\label{sec:segment_dpa}

Local affinities \(\alpha_{ij}\) (Eq.~\ref{eq:affinity_score}) must be
converted into a \emph{monotonic, one-to-one} mapping between source and target
segments. Monotonicity is a reasonable assumption since segments in each utterance
are temporally ordered and non-overlapping, and cross-lingual alignments are expected
to preserve the relative order of spoken content.

\paragraph{Sparse scoring grid.}
We define a DP table \(G \in \mathbb{R}^{(N+1) \times (M+1)}\) with:
\[
  G_{0,0} = 0, \qquad
  G_{i,0} = G_{0,j} = -\infty \quad (i,j > 0),
\]
where \(-\infty\) is implemented as \(-10^9\) to remain compatible with
\texttt{float32}.  
We compute \(G_{i,j}\) \emph{only} for candidate pairs
\((i,j) \in \mathcal{C}\) supplied by PCP
(§\ref{sec:segment_pairing1}); thus, both time and memory are
\(O(|\mathcal{C}|)\). Back-pointers are stored for the same sparse set.

\paragraph{Recurrence.}
We use a Needleman–Wunsch-style recurrence with gap penalty \(\gamma\)~\cite{needleman1970}.

\[
  G_{i,j} = 
  \max\!\bigl(
    G_{i-1,j-1} + \alpha_{ij},\;
    G_{i-1,j} + \gamma,\;
    G_{i,j-1} + \gamma
  \bigr).
\]
We set \(\gamma = -0.5\), chosen from a sweep over
\(\gamma \in \{-1.0, -0.75, -0.5, -0.25\}\) that maximised dev-set
alignment \(F_1\). This value is approximately half the median
\(\alpha_{ij}\), balancing match rewards against penalties for
insertion and deletion.

\paragraph{Back-trace.}
Starting at \(G_{N,M}\) and following stored pointers to \(G_{0,0}\)
yields an in-order alignment path:
\[
  \bigl\langle
    (s_{x_{i_1}}, s_{y_{j_1}}),\dots,
    (s_{x_{i_K}}, s_{y_{j_K}})
  \bigr\rangle,
  \qquad
  K \le \widetilde{N},
\]
where \(\widetilde{N}\) is the number of real (non-padded) source
segments from §\ref{sec:segment_pairing1}.  
This path constitutes the final segment alignment used for transfer and
all downstream evaluation.
Algorithm~\ref{alg:seg_alignment} summarizes the full SPaDA alignment process, combining pause, prosody, and semantic constraints into a unified scoring and decoding procedure.

\begin{algorithm}[H]
\caption{Pause- and Prosody-Constrained Segment Alignment (SPaDA)}
\label{alg:seg_alignment}
\begin{algorithmic}[1]
\REQUIRE Source utterance $x$, target utterance $y$
\ENSURE Monotonic alignment $\mathcal{A}$ between source and target segments

\STATE \textbf{Segmentation:} Apply VAD-based segmentation (§\ref{sec:speech_segmentation}) to $x$ and $y$, yielding segment pools $\{s_{x_i}\}_{i=1}^N$ and $\{s_{y_j}\}_{j=1}^M$

\STATE \textbf{Padding:} Equalize segment counts by appending silence-only segments to the shorter pool (§\ref{sec:segment_pairing1})

\STATE \textbf{Statistics:} Compute average durations $\mu_x$, $\mu_y$ and pause deltas $\Delta$, $\mu_{D_O}$, $\mu_{D_F}$

\FOR{each source segment $s_{x_i}$}
    \STATE \textbf{Candidate Generation (PCP):} Identify feasible target segments $\mathcal{C}_{x_i}$ using silence-consistency heuristics (§\ref{sec:segment_pairing1})
    \FOR{each candidate $s_{y_j} \in \mathcal{C}_{x_i}$}
        \STATE \textbf{Silence Consistency:} Compute $S_{\text{consistency}}(s_{x_i}, s_{y_j})$ via Eq.~\ref{eq:consistency_score}
        \STATE \textbf{Prosody Weighting:} Compute $P_{\text{prosody}}(i, j)$ from rate synchrony (§\ref{sec:prosody_pairing3})
        \STATE \textbf{Semantic Similarity:} Compute cosine similarity $\cos(f_s(s_{x_i}), f_s(s_{y_j}))$ (§\ref{sec:contrastive_pairing})
        \STATE \textbf{Affinity Score:} Combine all cues via Eq.~\ref{eq:affinity_score} to yield $\alpha_{ij}$
    \ENDFOR
\ENDFOR

\STATE \textbf{Alignment Decoding:}
\IF{greedy strategy}
    \STATE Select $\arg\max_j \alpha_{ij}$ for each $s_{x_i}$ (§\ref{sec:greedy_selection})
\ELSE
    \STATE Run dynamic programming with recurrence on $\alpha_{ij}$ (§\ref{sec:segment_dpa})
\ENDIF

\RETURN Final alignment $\mathcal{A}$
\end{algorithmic}
\end{algorithm}

\subsection{Guided Translation}
\label{sec:guided_tr}

\begin{figure*}[ht]
    \centering
    \includegraphics[scale=0.44]{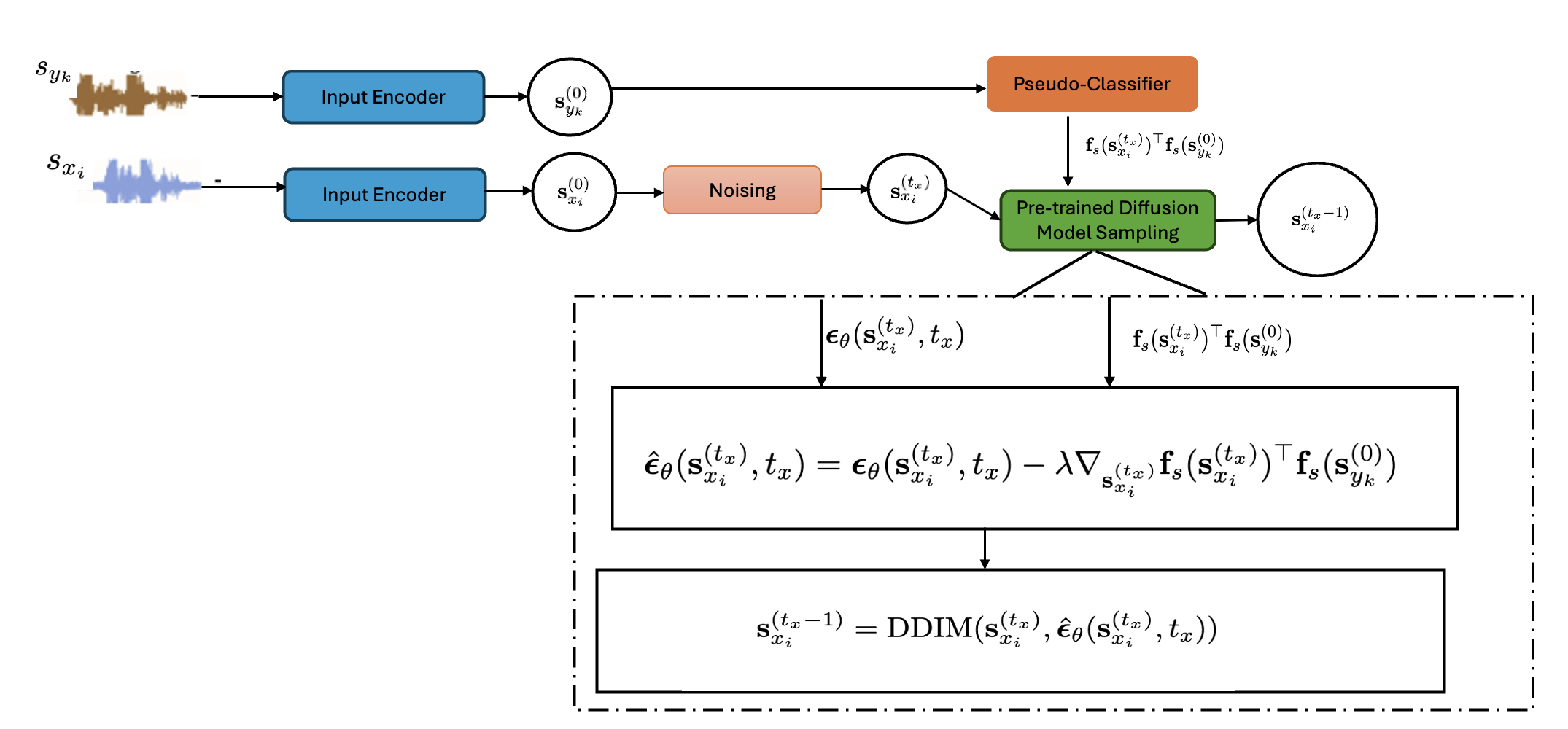}
    \caption{%
        Guided-diffusion pipeline for direct speech-to-speech translation (S2ST),  
        using $T\!=\!40$ DDIM iterations and guidance scale $\lambda\!=\!0.30$.  
        \textbf{(1)} An acoustic encoder $f_{\text{enc}}$ maps the \emph{source}
        segment $\mathbf{s}_{x_i}$ and a \emph{reference-voice} segment
        $\mathbf{s}_{y_k}$ to latent pseudo-spectrograms
        $\mathbf{s}_{x_i}^{(0)}$ and $\mathbf{s}_{y_k}^{(0)}$.  
        \textbf{(2)} $\mathbf{s}_{x_i}^{(0)}$ is corrupted to
        $\mathbf{s}_{x_i}^{(t)}$ at a random timestep
        $t\!\sim\!\mathcal{U}\{1,T\}$.  
        \textbf{(3)} A frozen semantic encoder $f_s$ (shared with
        §\ref{sec:contrastive_pairing}) measures cosine similarity between
        $\mathbf{s}_{x_i}^{(t)}$ and $\mathbf{s}_{y_k}^{(0)}$; its input-gradient
        steers the sample toward semantic faithfulness and the
        reference speaker’s voice characteristics.  
        \textbf{(4)} Each DDIM step adjusts the predicted noise with this
        gradient; the final denoised latent is vocoded to waveform.}
    \label{fig:guided_translation}
\end{figure*}

\noindent
The pipeline in Figure~\ref{fig:guided_translation} unfolds in four stages.

\paragraph{Step 1 – Latent encoding.}
We begin with a source segment $\mathbf{s}_{x_i}$ and a reference segment $\mathbf{s}_{y_k}$, each represented as $\mathbb{R}^{1 \times L}$. These are encoded by $f_{\text{enc}}$ (§\ref{sec:contrastive_pairing}):
\[
  \mathbf{s}_{x_i}^{(0)} = f_{\text{enc}}(\mathbf{s}_{x_i}), \qquad
  \mathbf{s}_{y_k}^{(0)} = f_{\text{enc}}(\mathbf{s}_{y_k}).
\]
These latents retain phonetic and coarse-prosodic information while removing the source speaker’s identity; the reference latent will later inject the target speaker’s voice characteristics.

\paragraph{Step 2 – Diffusion corruption.}
Sample $t\!\sim\!\mathcal{U}\{1,T\}$ and corrupt the source latent using forward noise:
\[
  \mathbf{s}_{x_i}^{(t)}
    = q\!\bigl(\mathbf{s}_{x_i}^{(0)},t\bigr)
    = \sqrt{\bar\alpha_t}\,\mathbf{s}_{x_i}^{(0)}
      + \sqrt{1-\bar\alpha_t}\,\boldsymbol\epsilon,
  \qquad \boldsymbol\epsilon\!\sim\!\mathcal{N}(\mathbf{0},\mathbf{I}),
\]
following the cosine noise schedule from~\cite{nichol2021improved}.

\paragraph{Step 3 – Gradient guidance.}
We compute cosine similarity between the noised source and reference latents using a frozen, unit-normalised encoder $f_s$:
\[
  g\!\bigl(\mathbf{s}_{x_i}^{(t)},\mathbf{s}_{y_k}^{(0)}\bigr)
    = f_s\!\bigl(\mathbf{s}_{x_i}^{(t)}\bigr)^{\!\top}
      f_s\!\bigl(\mathbf{s}_{y_k}^{(0)}\bigr) \in [-1,1],
\]
and use its input-gradient to guide the denoising:
\[
  \hat{\boldsymbol\epsilon}_\theta
  = \boldsymbol\epsilon_\theta\!\bigl(\mathbf{s}_{x_i}^{(t)},t\bigr)
    - \lambda\,
      \nabla_{\!\mathbf{s}_{x_i}^{(t)}}
      g\!\bigl(\mathbf{s}_{x_i}^{(t)},\mathbf{s}_{y_k}^{(0)}\bigr),
  \quad\;\lambda=0.30.
\]

\paragraph{Step 4 – DDIM sampling and synthesis.}
We iteratively apply:
\[
  \mathbf{s}_{x_i}^{(t-1)}
    = \text{DDIM}\!\bigl(
        \mathbf{s}_{x_i}^{(t)},\,
        \hat{\boldsymbol\epsilon}_\theta,\,
        t\bigr),\quad
  t=T,\dots,1,
\]
until reaching $\mathbf{s}_{x_i}^{(0)}$, which is then vocoded to waveform. The guidance preserves the source segment’s semantics while adapting to the reference speaker’s voice, enabling single-pass, speaker-consistent S2ST without text supervision.

\subsubsection{Unconditional Pre-trained Diffusion Sampling for S2ST}
\label{sec:pseudo_guidance}

Inspired by \textsc{GLIDE}~\cite{nichol2021glide}, which steers image generation using cosine similarity from a frozen text-image encoder (CLIP), we adopt a similar strategy for speech-to-speech translation (S2ST). Rather than relying on a trained classifier \( p(\mathbf{s}_{y_k}^{(0)}|\mathbf{s}_{x_i}^{(t)}) \) as in Eq.~\ref{eq:guide_w}, we guide an unconditional diffusion model using a pseudo-classifier—defined as the cosine similarity between segment embeddings from a frozen encoder. Specifically, the pre-trained model predicts noise \(\epsilon_\theta(\mathbf{s}_{x_i}^{(t)}, t)\), and we steer generation toward a semantically faithful output using a similarity-based guidance term:
\begin{equation}
\hat\epsilon_\theta\bigl(\mathbf{s}_{x_i}^{(t)},t\bigr)
  \;=\;
  \epsilon_\theta\bigl(\mathbf{s}_{x_i}^{(t)},t\bigr)
  \;-\;
  \lambda\,\sqrt{1-\bar\alpha_t}\,
  \nabla_{\mathbf{s}_{x_i}^{(t)}}\,
  \bigl[
     f_s\!\bigl(\mathbf{s}_{x_i}^{(t)}\bigr)^{\!\top}
     f_s\!\bigl(\mathbf{s}_{y_k}^{(0)}\bigr)
  \bigr],
\label{eq:guided_noise_revised}
\end{equation}

\noindent
where:
\begin{itemize}
  \item \( \epsilon_\theta(\cdot) \) is the noise estimate from the unconditional diffusion model;
  \item \( \bar\alpha_t = \prod_{j=1}^{t} \alpha_j \) is the cumulative noise schedule;
  \item \( f_s(\cdot) \) is a frozen, unit-normalized semantic segment encoder (§\ref{sec:contrastive_pairing});
  \item \( \lambda \) controls the strength of guidance.
\end{itemize}

The cosine similarity
\[
g\bigl(\mathbf{s}_{x_i}^{(t)}, \mathbf{s}_{y_k}^{(0)}\bigr) := f_s(\mathbf{s}_{x_i}^{(t)})^\top f_s(\mathbf{s}_{y_k}^{(0)})
\]
acts as a pseudo-classifier score. Though not trained as a classifier, it functions analogously: its gradient nudges the sample toward semantic alignment with the reference. This process occurs entirely externally—no conditioning input is passed to the diffusion model itself.

\vspace{5pt}
\paragraph{Clean-vs-noisy ablation.}
To test whether guidance quality depends on matching the diffusion timestep, we also evaluate a clean-guided variant where gradients are taken with respect to the clean latent:
\begin{equation}
\hat\epsilon_\theta\bigl(\mathbf{s}_{x_i}^{(t)},t\bigr)
  = \epsilon_\theta(\mathbf{s}_{x_i}^{(t)},t)
    - \lambda\,\sqrt{1-\bar\alpha_t}\;
      \nabla_{\mathbf{s}_{x_i}^{(0)}}
      \bigl[
        f_s(\mathbf{s}_{x_i}^{(0)})^{\!\top}
        f_s(\mathbf{s}_{y_k}^{(0)})
      \bigr].
\label{eq:ablation_guided_noise}
\end{equation}
This variant probes whether guidance remains effective when gradient information comes from a timestep-mismatched (clean) anchor.

\subsubsection{Conditional Pre-trained Diffusion Sampling for S2ST}
\label{sec:conditional_guidance}

Equations~\ref{eq:guided_noise_revised} and~\ref{eq:ablation_guided_noise} describe guidance in the unconditional case. We now extend this to the conditional setting, where the diffusion model explicitly conditions on the clean source segment \( \mathbf{s}_{y_k}^{(0)} \).

In this setting, the model predicts noise as:
\begin{equation}\label{eq:cond_guided1}
    \hat{\boldsymbol{\epsilon}}(\mathbf{s}_{x_i}^{(t_x)}|\mathbf{s}_{y_k}^{(0)}) =
    \boldsymbol{\epsilon}_\theta(\mathbf{s}_{x_i}^{(t_x)}, \mathbf{s}_{y_k}^{(0)}, t_x) -
    \lambda\sqrt{1 - \bar{\alpha}_{t_x}} \;
    \nabla_{\mathbf{s}_{x_i}^{(t_x)}}
    \left(
      f_s(\mathbf{s}_{x_i}^{(t_x)})^\top f_s(\mathbf{s}_{y_k}^{(0)})
    \right)
\end{equation}

Guidance is computed as a cosine similarity gradient between the current sample and the reference, using the same frozen segment encoder \( f_s \) as in §\ref{sec:pseudo_guidance}; this acts as a pseudo-classifier logit that nudges generation toward semantic alignment.

\vspace{5pt}
\paragraph{Clean-guided ablation.}
To assess whether the guidance signal benefits from timestep alignment, we substitute the clean embedding in place of the noisy one:
\begin{equation}\label{eq:cond_guided22}
    \hat{\boldsymbol{\epsilon}}(\mathbf{s}_{x_i}^{(t_x)}|\mathbf{s}_{y_k}^{(0)}) = 
    \boldsymbol{\epsilon}_\theta(\mathbf{s}_{x_i}^{(t_x)}, \mathbf{s}_{y_k}^{(0)}, t_x) -
    \lambda\sqrt{1 - \bar{\alpha}_{t_x}} \;
    \nabla_{\mathbf{s}_{x_i}^{(t_x)}}
    \left(
      f_s(\mathbf{s}_{x_i}^{(0)})^\top f_s(\mathbf{s}_{y_k}^{(0)})
    \right)
\end{equation}
This variant isolates whether using the clean (rather than timestep-matched) target embedding affects semantic consistency during generation.

\subsection{Unified Diffusion Pre-trained Model}

Instead of training separate unconditional and conditional diffusion models, we adopt a unified diffusion framework~\cite{bao2023one} that jointly models both marginal and conditional distributions over paired speech segments. This approach reduces training complexity and improves generalization across inference modes. Specifically, the model learns to approximate the following distributions:
   Marginals: \( q(\mathbf{s}_{x_i}) \), \( q(\mathbf{s}_{y_k}) \), conditional: \( q(\mathbf{s}_{x_i}|\mathbf{s}_{y_k}) \) and  joint: \( q(\mathbf{s}_{x_i}, \mathbf{s}_{y_k}) \). In the marginal case, the model predicts noise for a noised source segment \( \mathbf{s}_{x_i}^{(t_x)} \) without any conditioning:

\begin{equation}
    \mathbb{E}[\boldsymbol{\epsilon}^{x}| \mathbf{s}_{x_i}^{(t_x)}]
\end{equation}

For conditional and joint modeling, the noise prediction is conditioned on the clean or noised target segment:

\begin{equation}
    \mathbb{E}[\boldsymbol{\epsilon}^{x}|\mathbf{s}_{x_i}^{(t_x)}, \mathbf{s}_{y_k}^{(0)}], \qquad
    \mathbb{E}[\boldsymbol{\epsilon}^{x}, \boldsymbol{\epsilon}^{y}|\mathbf{s}_{x_i}^{(t_x)}, \mathbf{s}_{y_k}^{(t_y)}]
\end{equation}

Here, \( \boldsymbol{\epsilon}^{x}, \boldsymbol{\epsilon}^{y} \sim \mathcal{N}(0, I) \) are independent Gaussian noise vectors, and \( t_x, t_y \in \{1, \dots, T\} \) are diffusion timesteps.

The unified model is trained to predict concatenated noise vectors from paired noisy inputs:

\begin{equation}
    \mathbb{E}_{t_x, t_y, \boldsymbol{\epsilon}^{x}, \boldsymbol{\epsilon}^{y}} \left[
        \left\| 
            \boldsymbol{\epsilon}_\theta\bigl(\mathbf{s}_{x_i}^{(t_x)}, \mathbf{s}_{y_k}^{(t_y)}, t_x, t_y\bigr)
            - [\boldsymbol{\epsilon}^{x}, \boldsymbol{\epsilon}^{y}]
        \right\|^2
    \right]
\end{equation}

This objective enables the model to learn all three distributions using a single network. The unified model supports different inference modes by controlling \( t_y \):

\begin{itemize}
    \item Marginal sampling: Set \( t_y = T \), i.e., treat \( \mathbf{s}_{y_k}^{(t_y)} \) as fully noised. The model performs unconditional generation of \( \mathbf{s}_{x_i} \).
    \item Conditional sampling: Set \( t_y = 0 \), i.e., condition on a clean reference \( \mathbf{s}_{y_k}^{(0)} \). The model approximates \( q(\mathbf{s}_{x_i}|\mathbf{s}_{y_k}) \).
\end{itemize}

This unified design enables flexible switching between marginal, conditional, and joint generation using a single trained model, improving both training efficiency and inference generality.

\begin{figure*}[ht]
    \centering
    \includegraphics[scale=0.5]{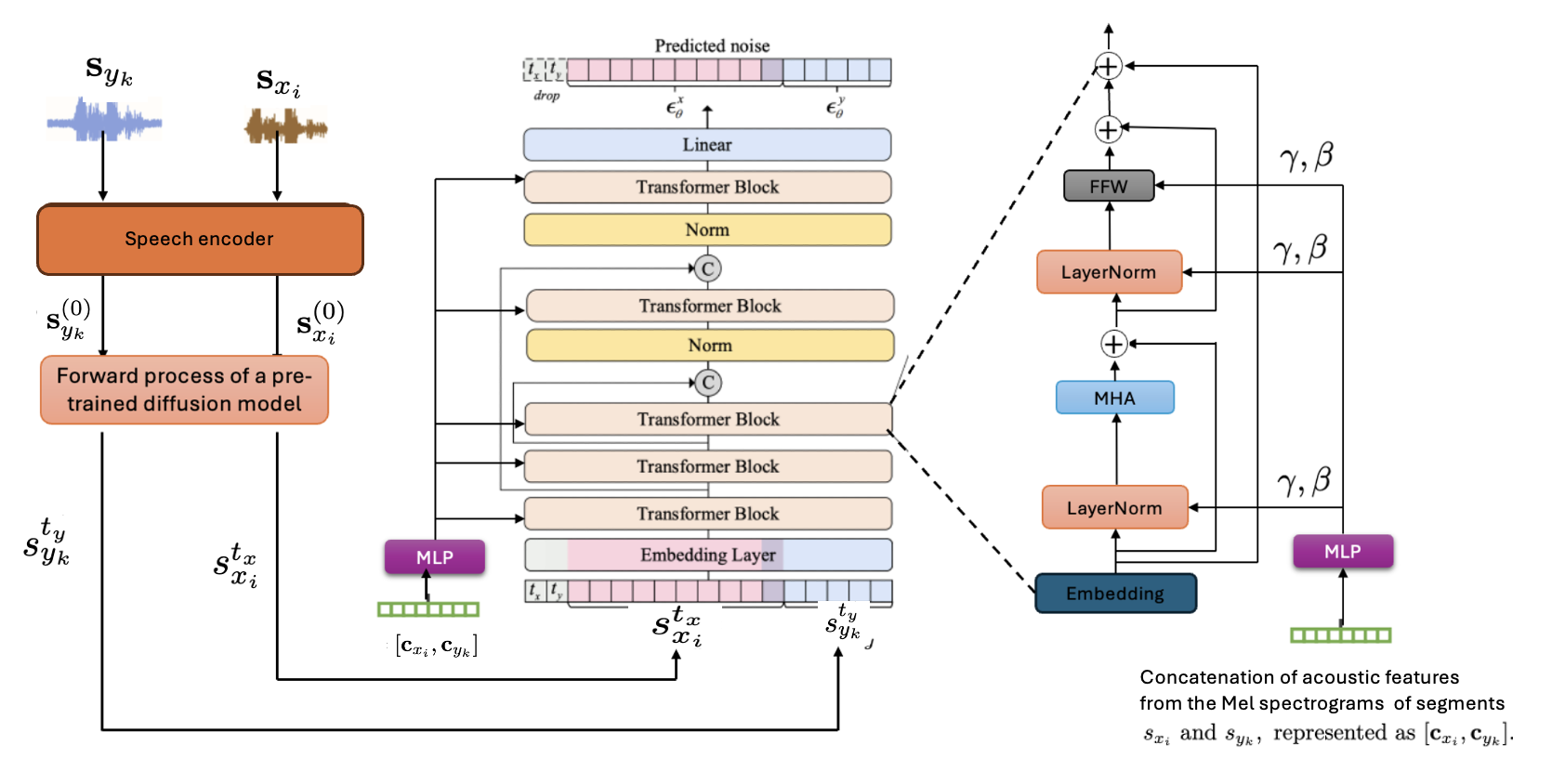}
    \caption{
        Unified Diffusion Model Architecture. Speech embeddings are perturbed with Gaussian noise and passed through Transformer blocks for noise prediction. Acoustic context, provided by Mel-spectrogram features, modulates intermediate representations via FiLM conditioning.
    }
    \label{fig:guided_translation2}
\end{figure*}

\subsubsection{Training and Sampling of the Unified Diffusion Model}
\label{sec:unified_diffusion}

Our unified diffusion model is trained once and reused for \emph{three} inference modes—marginal, conditional, and joint—by varying which of the paired speech segments is noised. Figure~\ref{fig:guided_translation2} illustrates the two-stage training process. Raw waveform segments \( \mathbf{s}_{x_i} \), \( \mathbf{s}_{y_k} \) are zero-padded to match the longest sequence in the batch and passed through a 1D convolutional encoder \( f_{\operatorname{enc}} \):
\[
  \mathbf{s}^{(0)}_{x_i} = f_{\operatorname{enc}}(\mathbf{s}_{x_i}),
  \qquad
  \mathbf{s}^{(0)}_{y_k} = f_{\operatorname{enc}}(\mathbf{s}_{y_k}),
\]
yielding latent spectrograms of dimension \( D = 256 \). We apply the standard cosine schedule from~\cite{nichol2021improved} to corrupt the latents with additive Gaussian noise. For diffusion steps \( t_x, t_y \sim \mathcal{U}\{1,\dots,T\} \)  the noised latents are:
\[
  \mathbf{s}^{(t_x)}_{x_i} = \sqrt{\bar\alpha_{t_x}}\,\mathbf{s}^{(0)}_{x_i} + \sqrt{1 - \bar\alpha_{t_x}}\,\boldsymbol\epsilon^{x},
  \qquad
  \boldsymbol\epsilon^{x} \sim \mathcal{N}(0, \mathbf{I}),
\]
and analogously for \( \mathbf{s}^{(t_y)}_{y_k} \). The paired noised latents \( (\mathbf{s}^{(t_x)}_{x_i}, \mathbf{s}^{(t_y)}_{y_k}) \) are fed into an 8-layer Transformer–UNet with FiLM-based conditioning. Each block contains multi-head attention (number of heads = 8), feed-forward layers, and residual connections. To inject acoustic context, we concatenate the Mel-spectrograms of both utterances:
\[
  \mathbf{c}_{\text{concat}} = [\mathbf{c}_{x_i}, \mathbf{c}_{y_k}] \in \mathbb{R}^{B \times (2L) \times C}.
\]
This is passed through a two-layer MLP:
\[
  \gamma, \beta = f_{\operatorname{MLP}}(\mathbf{c}_{\text{concat}}), \quad
  f_{\operatorname{MLP}}: \mathbb{R}^{B \times (2L) \times C} \to \mathbb{R}^{B \times L \times D},
\]
with hidden width \( H = 4D \) and GELU activations. The FiLM parameters are applied at each layer \( l \) as:
\[
  \text{FiLM}(\mathbf{h}^{(l)}) = \gamma \odot \mathbf{h}^{(l)} + \beta,
\]
where \( \mathbf{h}^{(l)} \in \mathbb{R}^{B \times L \times D} \) is the intermediate Transformer output and \( \odot \) denotes element-wise multiplication.

\paragraph{Mode Token Conditioning.}
A 3-way learned mode embedding \( \mathbf{z}_{\text{mode}} \in \mathbb{R}^{D} \), indicating \texttt{marginal}, \texttt{conditional}, or \texttt{joint}, is added to the diffusion timestep embedding before each Transformer block.

\paragraph{Training Objective.}
The model learns to predict the concatenated noise:
\[
\mathcal{L} = \mathbb{E}_{t_x, t_y, \boldsymbol{\epsilon}^{x}, \boldsymbol{\epsilon}^{y}} \left\| \boldsymbol{\epsilon}_\theta(\mathbf{s}^{(t_x)}_{x_i}, \mathbf{s}^{(t_y)}_{y_k}, t_x, t_y, \mathbf{z}_{\text{mode}}) - [\boldsymbol{\epsilon}^{x}, \boldsymbol{\epsilon}^{y}] \right\|_2^2,
\]
where \( [\boldsymbol{\epsilon}^{x}, \boldsymbol{\epsilon}^{y}] \in \mathbb{R}^{B \times L \times 2D} \) denotes channel-wise concatenation of the independent noise vectors.

\paragraph{Inference Modes.}
\begin{enumerate}[leftmargin=1.5em]
\item Marginal sampling: Noise both segments (\(t_x, t_y = T\)), set \( \mathbf{z}_{\text{mode}} = \texttt{marg} \), and denoise \( \mathbf{s}_x \) unconditionally.
\item Conditional sampling: Keep \( \mathbf{s}_{y_k}^{(0)} \) clean (i.e., \( t_y = 0 \)), set \( \mathbf{z}_{\text{mode}} = \texttt{cond} \), and denoise \( \mathbf{s}_x \) conditioned on the target.
\item Joint sampling: Noise both streams and set \( \mathbf{z}_{\text{mode}} = \texttt{joint} \) to synchronously denoise the pair.
\end{enumerate}

This architecture allows all inference regimes to be handled by a single model without retraining or architecture changes. Pseudocode is provided in (Algorithms 1–3).

\begin{algorithm}[H]
\caption{Training the Unified Diffusion Model}
\label{algo:train}
\begin{algorithmic}[1]
\REPEAT
    \STATE Sample a pair of latent speech segments: \\
    \hspace{10pt} \( \mathbf{s}_{x_i}^{(0)}, \mathbf{s}_{y_k}^{(0)} \sim q(\mathbf{s}_{x_i}, \mathbf{s}_{y_k}) \)
    \STATE Sample diffusion steps: \( t_x, t_y \sim \mathcal{U}\{1, \dots, T\} \)
    \STATE Sample noise vectors: \( \boldsymbol{\epsilon}^{x}, \boldsymbol{\epsilon}^{y} \sim \mathcal{N}(0, I) \)
    \STATE Compute noised latents: \\
    \hspace{10pt} \( \mathbf{s}_{x_i}^{(t_x)} = \sqrt{\bar{\alpha}_{t_x}} \mathbf{s}_{x_i}^{(0)} + \sqrt{1 - \bar{\alpha}_{t_x}} \boldsymbol{\epsilon}^{x} \)\\
    \hspace{10pt} \( \mathbf{s}_{y_k}^{(t_y)} = \sqrt{\bar{\alpha}_{t_y}} \mathbf{s}_{y_k}^{(0)} + \sqrt{1 - \bar{\alpha}_{t_y}} \boldsymbol{\epsilon}^{y} \)
    \STATE Select mode tag: \( \mathbf{z}_{\text{mode}} \in \{\text{marg}, \text{cond}, \text{joint}\} \)
    \STATE Predict noise: \\
    \hspace{10pt} \( \hat{\boldsymbol{\epsilon}} = \boldsymbol{\epsilon}_\theta(\mathbf{s}_{x_i}^{(t_x)}, \mathbf{s}_{y_k}^{(t_y)}, t_x, t_y, \mathbf{z}_{\text{mode}}) \)
    \STATE Compute and apply gradient: \\
    \hspace{10pt} \( \nabla_\theta \left\| \hat{\boldsymbol{\epsilon}} - [\boldsymbol{\epsilon}^{x}, \boldsymbol{\epsilon}^{y}] \right\|_2^2 \)
\UNTIL{convergence}
\end{algorithmic}
\end{algorithm}

\begin{algorithm}[H]
\caption{Unconditional Sampling (Marginal Generation)}
\label{algo:marginal}
\begin{algorithmic}[1]
\STATE Initialize \( \mathbf{s}_{x_i}^{(T)} \sim \mathcal{N}(0, I) \)
\FOR{\( t = T, T{-}1, \dots, 1 \)}
    \STATE Predict noise:
    \(
        \hat{\boldsymbol{\epsilon}}_\theta = \boldsymbol{\epsilon}_\theta(\mathbf{s}_{x_i}^{(t)}, \mathbf{s}_{y_k}^{(T)}, t, T, \mathbf{z}_{\text{mode}} = \text{marg})
    \)
    \STATE Update latent (DDIM-style):
    \(
        \mathbf{s}_{x_i}^{(t{-}1)} = \frac{1}{\sqrt{\bar{\alpha}_t}} \left( \mathbf{s}_{x_i}^{(t)} - \sqrt{1{-}\bar{\alpha}_t} \, \hat{\boldsymbol{\epsilon}}_\theta \right) \cdot \sqrt{\bar{\alpha}_{t{-}1}} + \sqrt{1{-}\bar{\alpha}_{t{-}1}} \, \boldsymbol{\epsilon}_{t-1}^{x}
    \)
\ENDFOR
\STATE \textbf{return} \( \mathbf{s}_{x_i}^{(0)} \)
\end{algorithmic}
\end{algorithm}

\begin{algorithm}[H]
\caption{Conditional Sampling (Guided Translation)}
\label{algo:conditional}
\begin{algorithmic}[1]
\STATE Initialize \( \mathbf{s}_{x_i}^{(T)} \sim \mathcal{N}(0, I) \)
\FOR{\( t = T, T{-}1, \dots, 1 \)}
    \STATE Predict guided noise:
    \(
        \hat{\boldsymbol{\epsilon}}_\theta = \boldsymbol{\epsilon}_\theta(\mathbf{s}_{x_i}^{(t)}, \mathbf{s}_{y_k}^{(0)}, t, 0, \mathbf{z}_{\text{mode}} = \text{cond}) 
        - \lambda \sqrt{1 - \bar{\alpha}_t} \nabla_{\mathbf{s}_{x_i}^{(t)}} \left( f_s(\mathbf{s}_{x_i}^{(t)}) \cdot f_s(\mathbf{s}_{y_k}^{(0)}) \right)
    \)
    \STATE Update latent (DDIM-style):
    \(
        \mathbf{s}_{x_i}^{(t{-}1)} = \frac{1}{\sqrt{\bar{\alpha}_t}} \left( \mathbf{s}_{x_i}^{(t)} - \sqrt{1{-}\bar{\alpha}_t} \, \hat{\boldsymbol{\epsilon}}_\theta \right) \cdot \sqrt{\bar{\alpha}_{t{-}1}}
    \)
\ENDFOR
\STATE \textbf{return} \( \mathbf{s}_{x_i}^{(0)} \)
\end{algorithmic}
\end{algorithm}

\section{Dataset, Segmentation, and Alignment Quality Evaluation}
This section introduces the dataset used in our experiments, describes the segmentation procedure (§~\ref{sec:speech_segmentation}), and evaluates the segment alignment strategies presented in §~\ref{sec:greedy_selection} and §~\ref{sec:segment_dpa}. Our evaluation combines: (i) statistical analysis of prosodic cues such as silence onsets and offsets; (ii) automatic scoring using BLEU based on ASR transcriptions; and (iii) human validation of segment pairs. Together, these analyses confirm the accuracy, consistency, and effectiveness of our segmentation and alignment pipeline.

\subsection{Dataset}
We collected a proprietary speech and text dataset from the Kenya Broadcasting Corporation (KBC), which operates 11 radio stations broadcasting in English and various Kenyan vernacular languages. News articles originally written in English were translated into vernacular languages and read by presenters at different times throughout the day. This process provided parallel news articles and their corresponding speech recordings, ensuring consistency across translations. The dataset features multiple newscasters, introducing diversity in speech patterns, accents, and vocal styles. This variation enhances its applicability to speech recognition and S2ST by capturing real-world speaker variability. We focused on 7 pm news bulletins, as they provide the most comprehensive daily broadcasts. The dataset spans news bulletins from 2018 to 2023. To ensure consistency, we removed advertisements and other non-news content. From the 11 available languages, we selected five widely spoken ones: Swahili, Luo, Kikuyu, Nandi, and English. These languages were chosen for their prevalence and cultural significance, representing major linguistic groups in Kenya. Table~\ref{tab:example} summarizes the dataset.

\begin{table}[ht]
    \centering
    \caption{Summary of the speech dataset used for evaluation.}
    \scalebox{0.7}{
    \begin{tabular}{cccc}
        \hline
        \textbf{Language} & \textbf{Phylum} & \textbf{\makecell{No. of \\News Bulletins}} & \textbf{\makecell{Total Speech Length (h)}} \\
        \hline
        Swahili & Niger-Congo & 2190 & 1353 \\
        Luo & Nilo-Saharan & 2190 & 1284 \\
        Kikuyu & Niger-Congo & 2190 & 1304 \\
        Nandi & Nilo-Saharan & 2190 & 1256 \\
        English & - & 2190 & 1206 \\
        \hline
    \end{tabular}}
    \label{tab:example}
\end{table}

This dataset serves as a rich resource for studying linguistic diversity and translation consistency across languages from distinct language families (Niger-Congo, Nilo-Saharan, and English as a global language). The pre-processing steps ensured semantic fidelity while eliminating inconsistencies such as advertisements. Additionally, the presence of multiple speakers per language enhances realism, making it well-suited for automatic speech recognition (ASR), speech synthesis, and S2ST tasks. 

\subsection{Segment Generation}
We segmented the news bulletins in the five target languages using the method described in §~\ref{sec:speech_segmentation}. Specifically, we applied an energy-based voice activity detector (VAD) with 0.2\,s padding, retaining only segments between 3 and 20 seconds in duration. Table~\ref{tab:segment_generation} reports the average segment length, median, inter-quartile range (IQR: 25th–75th percentile), and total number of retained segments per language.

\begin{table}[ht]
    \centering
    \caption{Segment generation statistics. Segment count reflects retained acoustic segments after VAD and length filtering; durations exclude long silences.}
    \scalebox{0.85}{\begin{tabular}{lcccc}
        \toprule
        \textbf{Language} & \textbf{Avg. Length (s)} & \textbf{Median (s)} & \textbf{IQR (s)} & \textbf{Total Segments} \\
        \midrule
        English & 15.0 & 14.7 & 6.7 & 665,578 \\
        Swahili & 15.2 & 15.0 & 6.1 & 638,944 \\
        Kikuyu  & 15.6 & 15.3 & 5.9 & 643,001 \\
        Luo     & 16.5 & 16.2 & 6.4 & 611,342 \\
        Nandi   & 17.1 & 16.8 & 6.5 & 599,456 \\
        \bottomrule
    \end{tabular}}
    \label{tab:segment_generation}
\end{table}

To assess whether linguistic phylum correlates with segment-length regularity, we compare intra- and inter-phylum distributions. Qualitatively, within-phylum pairs—e.g., Luo–Nandi (Nilo-Saharan)—exhibit closer average durations (0.6\,s difference) than cross-phylum pairs such as Luo–Kikuyu (0.9\,s difference). Despite the broad 3–20\,s segmentation window, the IQR values are narrowly distributed across all five languages (5.9–6.7\,s), indicating stable segmentation patterns even across typologically diverse language groups.

Moreover, languages within the same phylum exhibit closely aligned IQRs—Luo and Nandi (6.4\,s and 6.5\,s, Nilo-Saharan), and Kikuyu and Swahili (5.9\,s and 6.1\,s, Niger-Congo)—suggesting that shared phonotactic and prosodic features influence pause structure and segment regularity. In contrast, English shows a slightly broader IQR (6.7\,s), consistent with its role as a cross-phylum control and its broader stylistic and prosodic variation in broadcast speech.

While these patterns are suggestive, mean and IQR comparisons alone do not capture full distributional similarity. We therefore test the hypothesis that segment length distributions are more consistent within phyla than across them using a \emph{bootstrap resampling} strategy. Since each language yields approximately $6\times10^5$ segments, direct $p$-value testing would overstate significance. To mitigate this, we sample $10{,}000$ segments per language per pair and repeat this process over 1{,}000 replicates. On each replicate, we apply:

\begin{itemize}
    \item \textbf{Welch’s $t$-test} (mean difference, robust to variance heterogeneity)
    \item \textbf{Mann–Whitney $U$ test} (distributional shift, non-parametric)
\end{itemize}

We report median $p$-values across replicates and the absolute Cohen’s $d$ as a standardized effect size. Bonferroni correction is applied for four primary comparisons ($\alpha_\text{corr}=0.0125$). The within-phylum $p$-values (0.29–0.41) remain above this threshold.

\begin{table}[ht]
    \centering
    \caption{Distributional similarity tests across languages. Median $p$-values and absolute Cohen’s $d$ across 1\,000 bootstrap replicates ($n = 10{,}000$ per language). Language phylum affiliations are listed in Table~\ref{tab:example}.}
    \scalebox{0.9}{
    \begin{tabular}{lccc}
        \toprule
        \textbf{Pair} & \textbf{Median $p_\text{Welch}$} & \textbf{Median $p_U$} & \textbf{Median $|d|$} \\
        \midrule
        Luo–Nandi        & 0.34 & 0.29 & 0.07 \\
        Kikuyu–Swahili   & 0.41 & 0.38 & 0.05 \\
        \midrule
        Luo–Kikuyu       & $4.8\times10^{-4}$ & $6.1\times10^{-4}$ & 0.23 \\
        Nandi–Swahili    & $<10^{-6}$ & $<10^{-6}$ & 0.31 \\
        \midrule
        English–Luo      & $<10^{-6}$ & $<10^{-6}$ & 0.44 \\
        English–Swahili  & $<10^{-6}$ & $<10^{-6}$ & 0.37 \\
        \bottomrule
    \end{tabular}}
    \label{tab:segment_stats}
\end{table}

Within-phylum comparisons do not reach significance after correction and exhibit negligible effect sizes ($|d| \leq 0.07$). In contrast, cross-phylum pairs are consistently significant ($p < 0.001$) with small-to-moderate effects ($|d| = 0.23$–$0.31$). English–vernacular pairs show even stronger divergence, reinforcing the pattern. The results suggest that segment length distributions are statistically more consistent within phyla than across them. This may reflect underlying phonotactic and prosodic structure that subtly biases the VAD model's pause detection. Although absolute differences are small ($\leq6\%$ of the mean segment length), the analysis reveals a statistically grounded and practically meaningful trend: languages within the same phylum yield more uniform acoustic segmentations, validating our segmentation strategy for downstream ASR and speech-to-speech translation.
\subsection{Evaluation of Silence Timing Consistency and Correlation}
\label{sec:analysis}

To complement the segment duration analysis, we conduct a fine-grained evaluation of pause alignment across language pairs. Specifically, we assess the similarity of silence onsets and offsets using two metrics:

\begin{itemize}
    \item \textbf{Mean absolute difference and standard deviation (SD)} of pause timings — capturing the average temporal gap and variability across aligned segments.
    \item \textbf{Pearson correlation coefficient (PCC)} — quantifying the linear consistency of silence placement across paired utterances.
\end{itemize}

Silence onsets and offsets are measured relative to the start and end of each aligned sentence-level utterance, derived via alignment in (§~\ref{sec:segment_pairing1}). While SD reflects intra-pair variability, PCC captures whether silence timings are proportionally aligned—even if absolute shifts exist. Table~\ref{tab:silence_timing} presents results for ten language pairs based on 70,000 aligned segments per comparison.

\begin{table}[htbp]
\centering
\caption{Silence timing consistency across language pairs. All values are based on 70k aligned segment pairs per comparison.}
\label{tab:silence_timing}
\scalebox{0.7}{
\begin{tabular}{lcccccc}
\toprule
\textbf{Language Pair} & \textbf{Mean onset diff. [s]} & \textbf{SD onset [s]} & \textbf{Mean offset diff. [s]} & \textbf{SD offset [s]} & \textbf{PCC onset} & \textbf{PCC offset}  \\
\midrule
\multicolumn{7}{l}{\textit{Same-phylum pairs}} \\
Luo–Nandi (Nilo-Saharan)        & 0.407 & 0.289 & 0.439 & 0.249 & 0.551 & 0.475 \\
Kikuyu–Swahili (Niger-Congo)    & 0.446 & 0.261 & 0.471 & 0.281 & 0.475 & 0.463 \\
\midrule
\multicolumn{7}{l}{\textit{Cross-phylum pairs (sorted by onset diff.)}} \\
Nandi–Swahili                   & 0.584 & 0.492 & 0.725 & 0.633 & 0.165 & 0.143 \\
Luo–English                     & 0.629 & 0.524 & 0.609 & 0.626 & 0.133 & 0.169 \\
Swahili–English                 & 0.633 & 0.537 & 0.724 & 0.658 & 0.126 & 0.172 \\
Luo–Swahili                     & 0.642 & 0.519 & 0.765 & 0.719 & 0.154 & 0.195 \\
Kikuyu–English                  & 0.657 & 0.528 & 0.687 & 0.632 & 0.137 & 0.162 \\
Luo–Kikuyu                      & 0.683 & 0.664 & 0.676 & 0.649 & 0.169 & 0.132 \\
Nandi–Kikuyu                    & 0.722 & 0.565 & 0.566 & 0.523 & 0.143 & 0.148 \\
Nandi–English                   & 0.722 & 0.552 & 0.647 & 0.626 & 0.123 & 0.149 \\
\bottomrule
\end{tabular}
}
\end{table}

Same-phylum pairs Luo–Nandi and Kikuyu–Swahili exhibit lower mean and SD values for both onset and offset silence differences (e.g., mean onset differences: 0.41–0.45\,s), along with moderate PCCs (0.47–0.55). Given the large sample size, these correlations reflect substantial linear synchrony in pause structure across related languages. In contrast, \textit{cross-phylum pairs} show markedly larger timing mismatches (mean onset differences up to 0.72\,s), higher variability (SDs up to 0.66\,s), and weak correlations (PCC~$<$~0.20). For instance, Luo–Kikuyu and Nandi–Kikuyu display near-zero correlations, despite some lexical overlap, indicating deeper divergences in prosodic phrasing and pause structure.

To quantify these patterns, we perform a \textit{two-sided bootstrap test} over 1,000 replicates, resampling 10,000 segments per pair using seeds 0–999. The difference in mean onset timing between same- and cross-phylum groups yields an average $\Delta = 0.27$\,s with 95\% CI [0.22, 0.31], significant at $p < 0.01$.   These findings support the segment duration trends: languages within the same phylum show both tighter segment-length distributions and stronger pause alignment. This confirms that shared phonotactic and prosodic features influence acoustic segmentation, and it justifies the \emph{higher silence weight} (\(\lambda_s\)) used for within-phylum alignments in (Eq.~\ref{eq:affinity_score}).

\subsection{Segment-Pairing Variants}
\label{sec:pairing_variants}

To isolate the contribution of each affinity cue, we contrast three
\emph{greedy} methods (§\ref{sec:greedy_selection}) with the globally
optimal dynamic-programming aligner, \textsc{SPaDA}. Greedy methods rank
candidates locally by a scalar score and select the best match per segment,
while \textsc{SPaDA} enforces monotonic one-to-one alignment globally. Table \ref{tab:pair_counts_new} reports the number of aligned segment pairs
(in thousands) for each language combination. Percentages indicate the
relative increase in coverage over \textsc{SPaDA}.

\begin{table}[ht]
\centering
\caption{Aligned segment-pair counts (×10\textsuperscript{3}); $\;\%$ = relative gain over \textsc{SPaDA}.}
\label{tab:pair_counts_new}
\scalebox{0.88}{
\begin{tabular}{lrrrr}
\toprule
\textbf{Language pair} &
\textsc{PCP-G} &
\textsc{PCP+PW-G} &
\textsc{PCP+PW+SW-G} &
\textsc{SPaDA} \\ \midrule
\multicolumn{5}{l}{\emph{Within-phylum}}\\
Kikuyu–Swahili & 630\,(+5.0\%) & 618\,(+3.0\%) & 612\,(+2.0\%) & 600 \\
Luo–Nandi      & 609\,(+5.0\%) & 597\,(+2.9\%) & 592\,(+2.1\%) & 580 \\ \midrule
\multicolumn{5}{l}{\emph{Cross-phylum}}\\
Kikuyu–English & 622\,(+10.1\%) & 593\,(+5.0\%) & 582\,(+3.0\%) & 565 \\
Luo–Kikuyu     & 611\,(+10.1\%) & 583\,(+5.0\%) & 571\,(+2.9\%) & 555 \\
Luo–Swahili    & 605\,(+10.0\%) & 577\,(+4.9\%) & 566\,(+2.9\%) & 550 \\
Luo–English    & 600\,(+10.1\%) & 572\,(+5.0\%) & 560\,(+2.8\%) & 545 \\
Nandi–English  & 594\,(+10.0\%) & 567\,(+5.0\%) & 556\,(+3.0\%) & 540 \\
Nandi–Kikuyu   & 605\,(+10.0\%) & 577\,(+4.9\%) & 566\,(+2.9\%) & 550 \\
Nandi–Swahili  & 599\,(+9.9\%)  & 572\,(+5.0\%) & 560\,(+2.8\%) & 545 \\
Swahili–English& 646\,(+10.4\%) & 614\,(+5.0\%) & 602\,(+2.9\%) & 585 \\
\bottomrule
\end{tabular}}
\end{table}

For every language pair, we observe the same pattern:
\(
\textsc{PCP-G}> \textsc{PCP+PW-G}>\textsc{PCP+PW+SW-G}>\textsc{SPaDA},
\)
indicating that each additional cue progressively filters
out silence-induced over-alignments.

\paragraph{Within-phylum results.}
For related language pairs ( Kikuyu–Swahili and Luo–Nandi)
greedy coverage exceeds \textsc{SPaDA} by only 2–5\%, indicating that silence structure  provides a  strong basis for reliable alignment signal when
prosodic systems are similar.

\paragraph{Cross-phylum results.}
In cross-phylum settings, silence-only matching yields inflated pair counts
(9–11\% over \textsc{SPaDA}), often producing spurious alignments.
Articulation-rate filtering (\textsc{+PW}) prunes roughly half of these,
while semantic similarity (\textsc{+SW}) eliminates a further 40\% of
the remaining false positives. Manual inspection reveals that many
discarded pairs had incompatible lexical content despite similar silences.

\paragraph{Dynamic programming.}
\textsc{SPaDA} enforces global monotonicity while retaining
85–97\% of the pairs identified by \textsc{PCP-G}, primarily discarding
spurious or overlapping segments (2–10\% loss). This validates our weighting strategy:
silence cues dominate in structurally similar languages, while prosody and semantics
are crucial for robustness across more divergent pairs. The DP step adds structural
consistency with minimal recall loss.

\subsection{Data Split}
\label{sec:data}

For each language pair and alignment method (\textsc{PCP-G}, \textsc{PCP+PW-G}, \textsc{PCP+PW+SW-G}, and \textsc{SPaDA}), we partition the resulting segment pairs into 70\% training, 10\% development, and 20\% test splits. Since splitting is performed \textit{after} alignment, each method produces its own independently constructed 70/10/20 partition.

\subsection{Segment-Encoder Training}
\label{sec:segenc}

We pre-train a segment encoder \( f_s \) to map variable-length speech segments into a shared semantic space using a contrastive learning objective. As described in (§\ref{sec:contrastive_pairing}), the model is trained using same-utterance segments as positives, following COLA~\cite{saeed2021cola}, with all other in-batch segments treated as negatives. This setup encourages the encoder to learn general-purpose, language-agnostic representations that reflect semantic similarity without requiring parallel supervision.

Training is conducted on the pooled training split (§\ref{sec:data}), spanning all five languages. Segments are zero-padded to match the longest segment in each batch (up to a 20-second cap), with attention masks applied accordingly. The encoder adopts a two-stage architecture (§\ref{sec:contrastive_pairing}): a 1D convolutional front-end, followed by an EfficientNet-B0 backbone and a projection head used only during training. We optimize for 1M steps on an NVIDIA A100 GPU using AdamW (\( \beta_1 = 0.9 \), \( \beta_2 = 0.999 \), \( \epsilon = 10^{-8} \), weight decay = \(10^{-2}\)). We use a batch size of 512 and an initial learning rate of \(10^{-4}\), decayed via cosine annealing. The contrastive temperature is set to \( \tau = 0.07 \), following SimCLR best practices.

\subsection{Segment Alignment Quality Evaluation}
\label{sec:eval1}

We assess segment alignment quality using indirect but scalable metrics based on ASR and MT outputs. This proxy evaluation allows for large-scale assessment of semantic fidelity without requiring manual annotation. As reference transcriptions, we use the original news articles read aloud during broadcast (§~\ref{sec:data}), professionally translated from English into Swahili, Luo, Kikuyu, and Nandi. We manually aligned semantically equivalent sentences across these languages to form 10 bilingual corpora (Table~\ref{tab:mt_models}), which serve both MT training and segment evaluation.

\paragraph{MT systems.} We trained Transformer-base models for each language pair using 32K joint BPE tokens for 100K steps. BLEU scores were computed with SacreBLEU~\cite{post2018call} (v2.5.1) using standardized settings.\footnote{\texttt{nrefs=1\,|\,case=mixed\,|\,tok=13a\,|\,smooth=exp\,|\,version=2.5.1}}

\paragraph{ASR systems.} For Luo, Nandi, and Kikuyu, we trained Squeezeformer~\cite{kim2022squeezeformer} models from scratch. For Swahili and English, we fine-tuned Whisper-small~\cite{radford2023robust}. Table~\ref{tab:asr_wer} reports WERs on held-out test sets.

\begin{table}[t]
    \centering
    \caption{ASR performance across languages (WER, \%).}
    \label{tab:asr_wer}
    \scalebox{0.75}{
    \begin{tabular}{cc}
        \toprule
        \textbf{Language} & \textbf{WER (\%)} \\ \midrule
        Nandi & 13.6 \\
        Luo & 14.2 \\
        Kikuyu & 14.4 \\
        Swahili & 9.8 \\
        English & 5.3 \\
        \bottomrule
    \end{tabular}}
\end{table}

\begin{table}[t]
    \centering
    \caption{Parallel corpora used for MT training and BLEU evaluation.}
    \label{tab:mt_models}
    \scalebox{0.75}{
    \begin{tabular}{ccc}
        \toprule
        \textbf{Language Pair} & \textbf{Paired Sentences} & \textbf{BLEU Score} \\ \midrule
        Luo–Nandi & 1.76M & 32.2 \\
        Luo–Kikuyu & 1.18M & 31.8 \\
        Nandi–Kikuyu & 1.32M & 27.3 \\
        Kikuyu–Swahili & 1.29M & 30.4 \\
        Kikuyu–English & 1.71M & 24.9 \\
        Swahili–English & 1.52M & 25.4 \\
        Luo–Swahili & 1.43M & 27.4 \\
        Luo–English & 1.34M & 28.1 \\ 
        Nandi–Swahili & 1.44M & 27.1 \\
        Nandi–English & 1.37M & 28.6 \\
        \bottomrule
    \end{tabular}}
\end{table}

We introduce three automatic protocols for assessing the \emph{semantic alignment} of source–target segment pairs \((\mathbf{s}_{x_i}, \mathbf{s}_{y_k})\), along with a consistency check to evaluate their agreement. All protocols rely on three components: (i) ASR systems for the source and target languages, (ii) a high-quality MT system for translating source to target, and (iii) a set of clean target-language reference sentences \( \mathcal{C}_y \), drawn from the original news article.

\subsection{Automatic Scoring Alternatives}
\label{sec:scoring_variants}

Large-scale tuning is impractical with human judgment alone, so we design
three fully automatic, reference-light BLEU variants. These differ in how both
the hypothesis (source-derived) and reference (target-derived) sentences are
constructed. All variants use LaBSE~\cite{feng2022labse}, a multilingual
sentence encoder trained on 109 languages with strong zero-shot performance,
to support semantic retrieval in low-resource settings.

We define a cross-lingual retrieval operator:
\[
  \operatorname{Search}_{\mathcal C_y}(t)\;=\;
  \arg\max_{r\in\mathcal C_y}
  \cos\!\bigl(\operatorname{LaBSE}(t),\operatorname{LaBSE}(r)\bigr),
\]
which selects the closest reference sentence \( r \) from a corpus
\( \mathcal{C}_y \) of clean target-language sentences.

\vspace{2pt}
\begin{enumerate}[leftmargin=1.7em]

\item \textbf{M1 (\textsc{Cascade-BLEU})} 
      \[
        \operatorname{BLEU}\bigl(
          \text{MT}(\text{ASR}(\mathbf{s}_{x_i})),
          \text{ASR}(\mathbf{s}_{y_k})
      \bigr).
      \]
      Fully automatic and symmetric, but includes compounded ASR and MT noise.
      Human validation appears in §\ref{sec:mapquality}.

\item \textbf{M2 (\textsc{Target-Search BLEU})}  
      \[
        t_{y_k}^{*} = \operatorname{Search}_{\mathcal C_y}
                      \bigl(\text{ASR}(\mathbf{s}_{y_k})\bigr),
        \qquad
        \operatorname{BLEU}\bigl(
          \text{MT}(\text{ASR}(\mathbf{s}_{x_i})),\; t_{y_k}^{*}
        \bigr).
      \]
      Replaces the noisy ASR target with its nearest human-written reference,
      reducing ASR noise on the reference side.

\item \textbf{M3 (\textsc{Dual-Search BLEU})}  
      \[
        t_{y_k}^{\dagger} =
          \operatorname{Search}_{\mathcal C_y}
          \bigl(\text{MT}(\text{ASR}(\mathbf{s}_{x_i}))\bigr),\quad
        t_{y_k}^{*} =
          \operatorname{Search}_{\mathcal C_y}
          \bigl(\text{ASR}(\mathbf{s}_{y_k})\bigr),
      \]
      \[
        \operatorname{BLEU}\bigl(t_{y_k}^{\dagger},\,t_{y_k}^{*}\bigr).
      \]
      Both hypothesis and reference are retrieved from clean sentences, removing
      MT and ASR noise, but making the score dependent on retrieval accuracy.

\end{enumerate}

\subsection{Evaluation of Segment Mapping Quality}
\label{sec:mapquality}
We evaluate four alignment strategies—\textsc{PCP-G}, \textsc{PCP+PW-G}, \textsc{PCP+PW+SW-G}, and \textsc{SPaDA}—across the ten language pairs listed in Table~\ref{tab:example}. Alignment quality is assessed using BLEU, with 95\% confidence intervals (CIs) computed via non-parametric bootstrap resampling~\citep{koehn2004statistical}. The evaluation proceeds as follows:

\begin{enumerate}
  \item For each language pair, sample 100 test sets, each containing 700 randomly selected segment pairs.
  \item For each test set:
    \begin{enumerate}[label=(\alph*)]
      \item Generate 1,000 bootstrap replicas by resampling the 700 pairs with replacement.
      \item For each replica, concatenate:
        \begin{itemize}
          \item The predicted target segments, based on the decoding protocol (M1, M2, or M3), and
          \item The corresponding reference segments, also method-dependent.
        \end{itemize}
      \item Compute BLEU on each concatenated prediction–reference pair. Record the mean score \( \bar{b} \), and extract the 2.5th and 97.5th percentiles to compute the 95\% CI: \( \bar{b} \pm \Delta \).
    \end{enumerate}
  \item Report the average BLEU \( \bar{b} \) across the 100 sets as the final score. CI half-widths \( \Delta \) are averaged to reflect cross-set variability.
\end{enumerate}

The results reported in Table~\ref{tab:bleu_four_methods} show that:
\begin{table}[ht]
\centering
\caption{Mean BLEU scores $\pm$ 95\% CIs for four alignment strategies under M1, M2 and M3}
\label{tab:bleu_four_methods}
\resizebox{\textwidth}{!}{%
\begin{tabular}{lccc|ccc|ccc|ccc}
\toprule
\multirow{2}{*}{\textbf{Language pair}} &
\multicolumn{3}{c|}{\textbf{PCP-Greedy}} &
\multicolumn{3}{c|}{\textbf{PCP+PW-Greedy}} &
\multicolumn{3}{c|}{\textbf{PCP+PW+SW-Greedy}} &
\multicolumn{3}{c}{\textbf{SPaDA}} \\
& M1 & M2 & M3 & M1 & M2 & M3 & M1 & M2 & M3 & M1 & M2 & M3 \\
\midrule
Luo–Nandi       & 30.1 $\pm$ 1.8 & 31.7 $\pm$ 3.0 & 31.3 $\pm$ 2.6 & 31.0 $\pm$ 1.6 & 33.2 $\pm$ 2.9 & 32.9 $\pm$ 2.5 & 32.0 $\pm$ 1.5 & 33.8 $\pm$ 2.8 & 34.0 $\pm$ 2.4 & 32.6 $\pm$ 1.7 & 35.7 $\pm$ 2.8 & 36.8 $\pm$ 2.5 \\
Luo–Kikuyu      & 24.2 $\pm$ 3.7 & 25.6 $\pm$ 4.5 & 25.2 $\pm$ 3.6 & 25.1 $\pm$ 3.8 & 26.7 $\pm$ 4.4 & 26.2 $\pm$ 3.7 & 25.8 $\pm$ 3.5 & 27.1 $\pm$ 4.2 & 27.5 $\pm$ 3.5 & 26.9 $\pm$ 3.6 & 28.1 $\pm$ 4.3 & 29.0 $\pm$ 3.1 \\
Nandi–Kikuyu    & 23.0 $\pm$ 3.0 & 25.1 $\pm$ 4.1 & 24.9 $\pm$ 3.3 & 23.8 $\pm$ 2.9 & 26.4 $\pm$ 4.0 & 26.0 $\pm$ 3.2 & 24.5 $\pm$ 3.1 & 27.0 $\pm$ 4.1 & 27.3 $\pm$ 3.4 & 25.6 $\pm$ 3.0 & 27.6 $\pm$ 4.0 & 26.4 $\pm$ 3.2 \\
Kikuyu–Swahili  & 31.1 $\pm$ 2.3 & 32.1 $\pm$ 3.8 & 31.8 $\pm$ 3.0 & 32.3 $\pm$ 2.2 & 33.0 $\pm$ 3.7 & 33.2 $\pm$ 3.0 & 33.1 $\pm$ 2.1 & 34.1 $\pm$ 3.6 & 34.3 $\pm$ 2.8 & 33.7 $\pm$ 2.2 & 34.5 $\pm$ 3.7 & 34.2 $\pm$ 2.9 \\
Kikuyu–English  & 23.1 $\pm$ 2.4 & 24.0 $\pm$ 2.9 & 23.9 $\pm$ 2.6 & 24.0 $\pm$ 2.3 & 25.2 $\pm$ 2.8 & 25.1 $\pm$ 2.5 & 25.0 $\pm$ 2.2 & 26.1 $\pm$ 2.7 & 26.3 $\pm$ 2.5 & 25.4 $\pm$ 2.2 & 26.6 $\pm$ 2.7 & 27.1 $\pm$ 2.5 \\
Swahili–English & 22.0 $\pm$ 3.0 & 23.7 $\pm$ 2.7 & 23.5 $\pm$ 2.5 & 23.4 $\pm$ 2.9 & 25.1 $\pm$ 2.6 & 25.0 $\pm$ 2.4 & 23.6 $\pm$ 2.7 & 25.7 $\pm$ 2.5 & 25.9 $\pm$ 2.3 & 23.9 $\pm$ 1.9 & 26.6 $\pm$ 2.5 & 26.4 $\pm$ 2.4 \\
Luo–Swahili     & 24.5 $\pm$ 2.8 & 24.7 $\pm$ 3.7 & 25.0 $\pm$ 3.1 & 25.3 $\pm$ 2.9 & 25.4 $\pm$ 3.6 & 25.6 $\pm$ 3.0 & 25.9 $\pm$ 2.8 & 26.3 $\pm$ 3.5 & 26.7 $\pm$ 3.1 & 26.7 $\pm$ 2.9 & 26.8 $\pm$ 3.6 & 27.2 $\pm$ 3.3 \\
Luo–English     & 22.2 $\pm$ 4.3 & 23.6 $\pm$ 2.3 & 23.3 $\pm$ 2.2 & 23.1 $\pm$ 4.2 & 25.3 $\pm$ 2.2 & 25.1 $\pm$ 2.1 & 23.8 $\pm$ 4.1 & 26.1 $\pm$ 2.1 & 26.4 $\pm$ 2.0 & 24.6 $\pm$ 4.3 & 26.8 $\pm$ 2.3 & 26.9 $\pm$ 2.1 \\
Nandi–Swahili   & 21.3 $\pm$ 2.6 & 21.2 $\pm$ 3.0 & 21.7 $\pm$ 2.8 & 22.6 $\pm$ 2.5 & 22.4 $\pm$ 2.9 & 22.7 $\pm$ 2.7 & 23.1 $\pm$ 2.4 & 23.5 $\pm$ 2.9 & 23.9 $\pm$ 2.7 & 24.1 $\pm$ 2.6 & 23.9 $\pm$ 3.0 & 24.4 $\pm$ 2.7 \\
Nandi–English   & 20.5 $\pm$ 4.5 & 22.4 $\pm$ 4.0 & 22.0 $\pm$ 3.6 & 21.7 $\pm$ 4.4 & 23.8 $\pm$ 3.9 & 23.6 $\pm$ 3.5 & 22.4 $\pm$ 4.2 & 24.5 $\pm$ 3.8 & 24.8 $\pm$ 3.6 & 23.2 $\pm$ 4.4 & 25.2 $\pm$ 4.0 & 25.6 $\pm$ 3.5 \\
\bottomrule
\end{tabular}}
\end{table}

\subparagraph*{(a) Model ranking.}
Across all ten language pairs and evaluation protocols, the model ranking is consistent:
\[
  \text{PCP-G} \;<\;
  \text{PCP+PW-G} \;<\;
  \text{PCP+PW+SW-G} \;<\;
  \text{SPaDA}.
\]

\subparagraph*{(b) Prosodic and semantic cues.}
\textit{Within-phylum pairs} (\textit{Luo–Nandi}, \textit{Kikuyu–Swahili}) perform relatively well even with silence-only alignment (PCP-G), achieving 30–32 BLEU under M1. Adding prosodic cues (PCP+PW-G) yields modest gains of +1.0 BLEU, while semantic weighting (PCP+PW+SW-G) adds a further +1.0 BLEU. Switching to global alignment with SPaDA improves scores by an additional +1.5 BLEU. However, the overall gains are incremental due to the high reliability of silence cues within related language pairs.

In contrast, \textit{cross-phylum pairs} (e.g., \textit{Nandi–English}, \textit{Swahili–English}) start at lower baselines (20–22 BLEU with PCP-G). Here, prosodic weighting provides a larger relative improvement (+1.2–1.6 BLEU), semantic embeddings add +1.0–1.3 BLEU, and SPaDA yields the most substantial boost—up to +2.5 BLEU under M3. These results underscore the importance of richer modeling when silence cues alone are unreliable, as in genealogically distant pairs.

\subparagraph*{(c) Evaluation protocols.}
M2 and M3 consistently raise absolute BLEU scores by approximately 1 point over M1 (the ASR–MT cascade), while preserving the model ranking. This suggests that reference cleaning via retrieval (Search) enhances measurement stability without distorting comparative performance.

\subsection{Binary Alignment Evaluation}
\label{sec:binary_eval}

While continuous BLEU scores provide a scalable proxy for semantic similarity, they do not indicate whether an individual segment pair is actually usable. We therefore reframe the task as binary classification and report precision, recall, and \(F_1\). Because BLEU distributions vary across languages and segment lengths, we avoid arbitrary thresholds and instead derive a data-driven cutoff \(\tau_{\text{BLEU}}\) through a small-scale human annotation study:

\begin{enumerate}[leftmargin=1.4em,itemsep=2pt]
\item For each language pair, we sample 40 segment pairs near three reference points—the 2.5th percentile, corpus mean, and corpus median—yielding \(10 \times 3 \times 40 = 1{,}200\) examples in total.
\item Two bilingual annotators label each pair as \textit{aligned} or \textit{misaligned}. Inter-annotator agreement is strong (\(\kappa = 0.79\), bootstrapped).
\item We select the BLEU value that maximizes \(F_1\) agreement with pooled human labels as the global decision threshold:
\[
\tau_{\text{BLEU}} = 22.7.
\]
Optimizing \(\tau\) separately per language changes macro-\(F_1\) by less than 0.2 points, so a single global threshold suffices.
\end{enumerate}

Given true positives (\(\mathrm{TP}\)), false positives (\(\mathrm{FP}\)), and false negatives (\(\mathrm{FN}\)), we compute:
\begin{equation}
\label{eq:prf_metrics}
\mathrm{Prec} = \frac{\mathrm{TP}}{\mathrm{TP} + \mathrm{FP}}, \quad
\mathrm{Rec} = \frac{\mathrm{TP}}{\mathrm{TP} + \mathrm{FN}}, \quad
F_1 = \frac{2\,\mathrm{Prec}\,\mathrm{Rec}}{\mathrm{Prec} + \mathrm{Rec}}.
\end{equation}

Table~\ref{tab:prf_binary_full} summarizes binary alignment performance for the two strongest systems—\textsc{PCP+PW+SW-G} and \textsc{SPaDA}—under all three evaluation protocols (M1, M2, and M3). We omit results for \textsc{PCP-G} and \textsc{PCP+PW-G}, as both consistently underperform PCP+PW+SW-G by 1.4–2.0 \(F_1\) points across all protocols. Our analysis thus focuses on the most competitive baselines.

\begin{table}[ht]
\centering
\caption{Binary alignment performance (Precision, Recall, \(F_1\)) under evaluation protocols M1, M2 and M3. Scores are averaged over 100 bootstrap folds.}
\label{tab:prf_binary_full}
\resizebox{\textwidth}{!}{%
\begin{tabular}{lcccccccccccccccccc}
\toprule
\multirow{3}{*}{\textbf{Language Pair}} 
& \multicolumn{9}{c}{\textbf{PCP+PW+SW-Greedy}} 
& \multicolumn{9}{c}{\textbf{SPaDA}} \\
\cmidrule(lr){2-10} \cmidrule(lr){11-19}
& \multicolumn{3}{c}{\textbf{M1}} 
& \multicolumn{3}{c}{\textbf{M2}} 
& \multicolumn{3}{c}{\textbf{M3}} 
& \multicolumn{3}{c}{\textbf{M1}} 
& \multicolumn{3}{c}{\textbf{M2}} 
& \multicolumn{3}{c}{\textbf{M3}} \\
\cmidrule(lr){2-4} \cmidrule(lr){5-7} \cmidrule(lr){8-10} 
\cmidrule(lr){11-13} \cmidrule(lr){14-16} \cmidrule(lr){17-19}
& \textbf{Prec} & \textbf{Rec} & \(\mathbf{F_1}\) 
& \textbf{Prec} & \textbf{Rec} & \(\mathbf{F_1}\) 
& \textbf{Prec} & \textbf{Rec} & \(\mathbf{F_1}\) 
& \textbf{Prec} & \textbf{Rec} & \(\mathbf{F_1}\) 
& \textbf{Prec} & \textbf{Rec} & \(\mathbf{F_1}\) 
& \textbf{Prec} & \textbf{Rec} & \(\mathbf{F_1}\) \\
\midrule
Luo--Nandi       & 74.0 & 72.0 & 72.9 & 75.9 & 73.9 & 74.9 & 77.3 & 75.1 & 76.2 & 76.8 & 74.3 & 75.5 & 78.5 & 76.2 & 77.3 & 79.7 & 77.9 & 78.8 \\
Kikuyu--Swahili  & 76.1 & 74.8 & 75.4 & 78.4 & 76.6 & 77.5 & 79.7 & 78.1 & 78.9 & 79.4 & 77.9 & 78.7 & 81.2 & 79.3 & 80.2 & 82.4 & 80.5 & 81.4 \\
Kikuyu--English  & 65.4 & 64.5 & 64.9 & 69.4 & 67.2 & 68.3 & 71.2 & 69.1 & 70.1 & 70.3 & 69.4 & 69.8 & 72.7 & 70.7 & 71.7 & 74.2 & 72.1 & 73.1 \\
Luo--Kikuyu      & 63.3 & 61.7 & 62.5 & 66.1 & 63.5 & 64.8 & 67.7 & 65.1 & 66.4 & 67.4 & 65.4 & 66.4 & 69.2 & 67.0 & 68.1 & 71.1 & 68.6 & 69.8 \\
Nandi--Kikuyu    & 60.7 & 59.5 & 60.1 & 64.0 & 62.1 & 63.0 & 65.9 & 63.9 & 64.9 & 65.0 & 63.1 & 64.0 & 67.5 & 65.1 & 66.3 & 69.3 & 67.0 & 68.1 \\
Luo--Swahili     & 68.7 & 66.9 & 67.8 & 71.4 & 69.1 & 70.2 & 73.0 & 70.9 & 71.9 & 73.5 & 71.3 & 72.4 & 75.7 & 73.2 & 74.4 & 77.5 & 74.8 & 76.1 \\
Swahili--English & 58.3 & 56.9 & 57.6 & 62.2 & 60.6 & 61.4 & 64.1 & 62.0 & 63.0 & 63.0 & 61.2 & 62.1 & 65.6 & 63.7 & 64.6 & 67.3 & 65.0 & 66.1 \\
Luo--English     & 60.4 & 58.6 & 59.5 & 64.0 & 61.2 & 62.6 & 66.2 & 63.7 & 64.9 & 66.0 & 64.0 & 65.0 & 68.3 & 66.0 & 67.1 & 70.0 & 67.8 & 68.9 \\
Nandi--Swahili   & 57.3 & 55.6 & 56.4 & 59.9 & 57.9 & 58.9 & 61.5 & 59.4 & 60.4 & 60.9 & 58.7 & 59.8 & 63.1 & 61.0 & 62.0 & 65.0 & 62.5 & 63.7 \\
Nandi--English   & 54.8 & 52.9 & 53.8 & 58.4 & 56.0 & 57.2 & 59.9 & 57.6 & 58.7 & 59.7 & 57.5 & 58.6 & 63.6 & 60.6 & 62.0 & 65.1 & 62.4 & 63.7 \\
\bottomrule
\end{tabular}}
\end{table}
Key Findings from Table ~\ref{tab:prf_binary_full} inlude:
\begin{enumerate}[leftmargin=1.3em,itemsep=4pt]
\item \textbf{Model ranking.}  
      \textsc{SPaDA} consistently outperforms the strongest greedy baseline
      (\textsc{PCP+PW+SW-G}) across all language pairs and evaluation
      protocols. Averaged over the ten pairs, the \(F_1\) gains are
      +4.0 (M1), +3.7 (M2), and +3.4 (M3), reflecting improvements in both precision and recall.

\item \textbf{Within– vs.\ cross-phylum.}  
      Within-phylum pairs outperform cross-phylum ones by approximately 12 \(F_1\) points, echoing the pause-correlation gap noted in §\ref{sec:analysis}. This suggests that richer modelling layers are especially helpful when pause cues are unreliable.

\item \textbf{Protocol sensitivity.}  
      Scores increase consistently as the reference becomes cleaner:
      \(\text{M1} < \text{M2} < \text{M3}\). 
\end{enumerate}

\subsection{Paired-Bootstrap Significance Test}
\label{sec:bootstrap_significance}

To verify that the three automatic protocols in
§\ref{sec:scoring_variants} produce genuinely different BLEU scores, we
conduct a paired bootstrap test following \citet{koehn2004statistical},
using the best aligner, \textsc{SPaDA}. For each language pair, we create 100 outer folds, each
containing 700 aligned segment pairs sampled without replacement. Within
each fold, we draw 1,000 bootstrap replicas (resampling with
replacement) and compute
\[
  \Delta = \text{BLEU}_{i} \;-\; \text{BLEU}_{j}.
\]
A fold is considered a \emph{significant win} for protocol \(i\) if
\(\Pr(\Delta > 0) \ge 0.95\) across its 1,000 replicas (one-tailed test).

\begin{table}[ht]
\centering
\caption{Paired-bootstrap comparison of BLEU protocols for
\textsc{SPaDA}. “Wins” = percentage of the 100 folds where the
95\,\% one-tailed criterion is met. Median $\Delta$ and 95\,\% confidence
intervals (CI) are computed over fold-level medians.}
\label{tab:bootstrap_bleu}
\small
\begin{tabular}{llccc}
\toprule
\textbf{Language Pair} & \textbf{Contrast}         &
\textbf{Wins (\%)} & \textbf{Median $\Delta$} & \textbf{95\,\% CI} \\ \midrule
\multirow{3}{*}{Luo → Kikuyu (cross)}
  & M2 $>$ M1 & 80 & +1.6 & [\,0.9, 2.4\,] \\
  & M3 $>$ M1 & 94 & +2.4 & [\,1.5, 3.3\,] \\
  & M3 $>$ M2 & 86 & +1.0 & [\,0.3, 1.9\,] \\ \midrule
\multirow{3}{*}{Luo → Nandi (within)}
  & M2 $>$ M1 & 73 & +1.1 & [\,0.4, 1.9\,] \\
  & M3 $>$ M1 & 91 & +2.0 & [\,1.2, 2.8\,] \\
  & M3 $>$ M2 & 79 & +0.9 & [\,0.2, 1.6\,] \\ \bottomrule
\end{tabular}
\end{table}

Table~\ref{tab:bootstrap_bleu} indicates the following findings:

In both a within-phylum (Luo → Nandi) and a cross-phylum (Luo → Kikuyu)
setting, the ordering
\[
  \text{M1} \;<\; \text{M2} \;<\; \text{M3}
\]
holds in the majority of folds, with median BLEU gaps of +1.0 to +2.4
points. Effects are more pronounced for
the cross-phylum pair, underscoring that anchoring evaluation to cleaner
textual references (M2, M3) produces a more reliable and discriminative
signal when language divergence increases. These results support our decision to report all three protocols: while
absolute scores vary with reference quality, the relative ranking of
alignment systems remains stable.

\subsection{Manual Segment-Pair Evaluation}
\label{sec:human_eval}
Automatic metrics are only proxies for semantic adequacy.  
We therefore conduct a human study over a \emph{stratified} sample of
1,000 unique segment pairs (100 per language pair) drawn from the
best-performing system, SPaDA.  
The sample is balanced for length (4–16 s) and BLEU decile to ensure 
both easy and difficult cases are represented. From 2,407 screened volunteers, we randomly selected
1,000 undergraduate speakers
(native Luo, Nandi, or Kikuyu; fluent in both Swahili and English).
Each annotator received 20 speech segments
(\(\approx10\) source–target pairs).\footnote{Pairs were assigned
\emph{with replacement}; on average, each pair was seen by
$\approx$10 annotators—providing redundancy.}
Compensation followed university ethics guidelines:  
KSh~200 (\(\approx\)\,\$1.50) for a \(\le\)5-minute task.

\paragraph{Two-stage labeling protocol.}
\begin{enumerate}[nosep,leftmargin=1.45em]
  \item \textbf{Transcription.}  
        Each source and target segment is independently transcribed
        by \(\ge\!2\) annotators fluent in the respective language.
  \item \textbf{Semantic rating.}  
        A separate bilingual duo reads both transcripts and assigns
        a 5-point similarity score  
        (5 = perfect paraphrase, 1 = unrelated).  
        Pairs with a \emph{mean} score \(>\!4\) are marked
        \emph{correctly aligned}.  
        Inter-rater agreement is strong (\(\kappa=0.82\)).
\end{enumerate}

\paragraph{Binary metrics.}
With \(C\) “correct’’ pairs, \(H{=}1,000\) usable pairs
(length/quality filter), and \(T{=}1,000\) total, precision, recall, and
\(F_1\) follow Eq.~\ref{eq:prf_metrics}.

\begin{table}[h]
\centering
\caption{Human \(F_1\) (\%) for SPaDA.}
\label{tab:human-f1}
\small
\setlength{\tabcolsep}{7pt}
\begin{tabular}{lS[table-format=2.1]}
\toprule
\textbf{Language pair} & {\(F_1\)} \\
\midrule
\multicolumn{2}{l}{\emph{Within-phylum}}\\
Luo–Nandi        & 77.0 \\
Kikuyu–Swahili   & 78.5 \\
\midrule
\multicolumn{2}{l}{\emph{Cross-phylum}}\\
Kikuyu–English   & 72.4 \\
Luo–Swahili      & 70.6 \\
Luo–Kikuyu       & 68.7 \\
Luo–English      & 68.0 \\
Nandi–Kikuyu     & 66.8 \\
Swahili–English  & 64.2 \\
Nandi–Swahili    & 63.1 \\
Nandi–English    & 61.3 \\
\bottomrule
\end{tabular}
\end{table}
Table~\ref{tab:human-f1} indicates the following findings:
\begin{itemize}[leftmargin=1.4em,itemsep=3pt]
  \item \textbf{Correlation with automatic scores.}  
        Human \(F_1\) values are 3–4 points lower than
        (M3), but rank the language pairs
        identically (\(\rho_{\text{Spearman}} = 0.92\)).
        Thus, automatic evaluation is slightly optimistic but
        reliably comparative.
  \item \textbf{Role of linguistic relatedness.}  
        The within-phylum advantage is substantial  
        (\(+8.3\) \(F_1\) on average), corroborating pause- and
        prosody-based findings in \S\ref{sec:analysis}
        and \S\ref{sec:binary_eval}.
  \item \textbf{Error taxonomy.}  
        Manual comments reveal two major sources:  
        (i) \textit{boundary errors} (e.g., late-start/early-end),
        and (ii) \textit{false friends}—semantically mismatched but
        acoustically similar phrases.  
        Both occur more in cross-phylum settings and are mitigated
        by prosodic cues in SPaDA.
\end{itemize}

\subsection{Effect of Segment Length on Mapping Quality}
\label{sec:length}

We investigate how segment length affects alignment quality by bucketing each bilingual pair \((s_{x_i},\,s_{y_k})\) based on its relative length and evaluating alignment performance using BLEU. 
Let \(l_x\) and \(l_y\) denote the average token lengths of languages \(x\) and \(y\), respectively (see Table~\ref{tab:segment_generation}). For each segment pair, we compute the midpoint length \(\bar{l} = \tfrac{1}{2}(l_x + l_y)\), and assign the pair to one of three categories:

\begin{itemize}[leftmargin=1.5em, nosep]
  \item \textbf{Short}: \(|s_{x_i}|,\,|s_{y_k}| < 0.7\,\bar{l}\)
  \item \textbf{Average}: \(0.7\,\bar{l} \le |s_{x_i}|,\,|s_{y_k}| \le 1.3\,\bar{l}\)
  \item \textbf{Long}: \(|s_{x_i}|,\,|s_{y_k}| > 1.3\,\bar{l}\)
\end{itemize}

(Thresholds 0.7 and 1.3 were chosen for symmetry; minor variations do not change the ranking.)

From the Luo–Nandi, Luo–Kikuyu, and Kikuyu–Swahili corpora, we sample \(n = 1{,}000\) disjoint segment pairs per bucket (total \(n = 3{,}000\)), ensuring no overlap with any prior test folds. Alignments are produced using SPaDA and evaluated under M3. For each bucket, we report the mean BLEU and 95\% confidence intervals from 1,000 bootstrap replicas. Figure~\ref{fig:length_impact} reveals two consistent trends across all three language pairs:

\begin{figure}[ht]
  \centering
  \includegraphics[width=\linewidth]{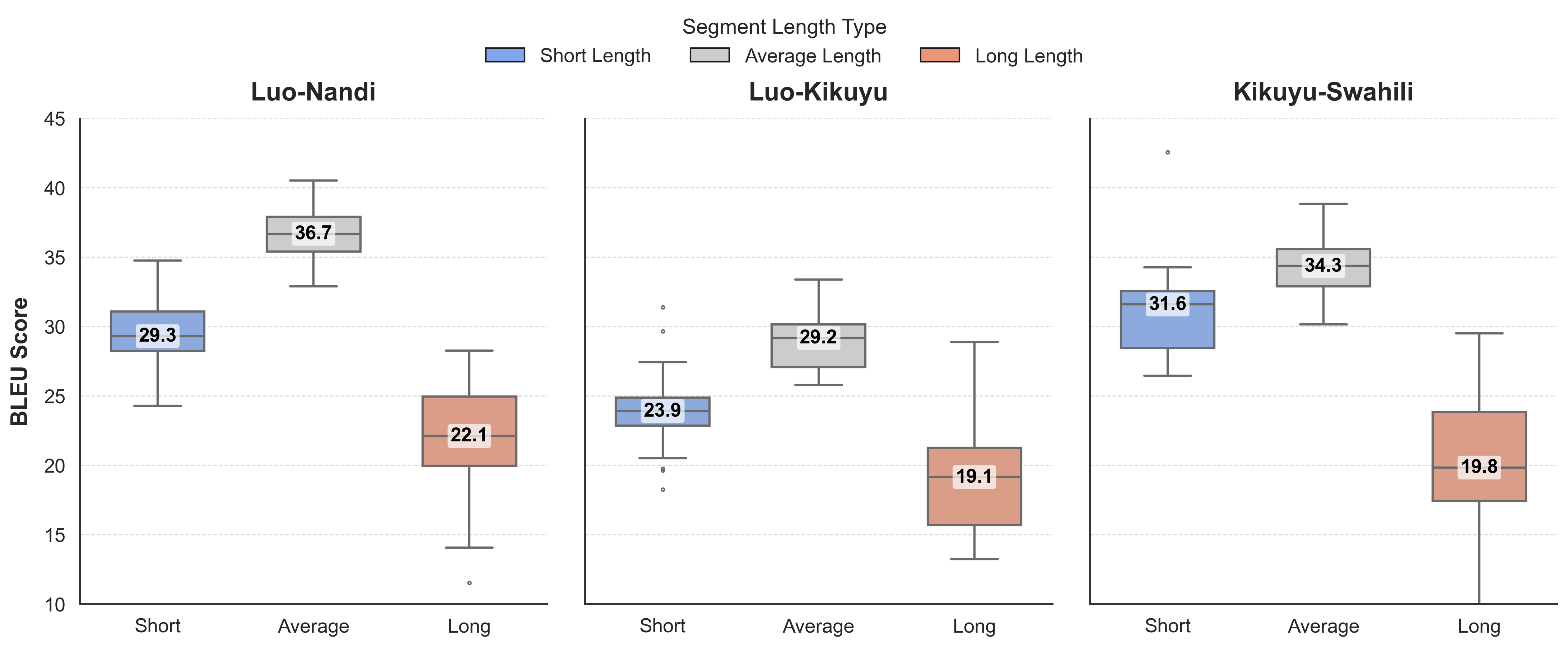}
  \caption{BLEU by segment-length bucket (\(n = 1{,}000\) per bucket).}
  \label{fig:length_impact}
\end{figure}

\begin{itemize}[leftmargin=1.4em,itemsep=4pt]
\item \textbf{Segment length impacts alignment quality.}  
      Average-length segments achieve the highest BLEU scores in every case:  
      36.7 (Luo–Nandi), 29.2 (Luo–Kikuyu), and 34.3 (Kikuyu–Swahili).

\item \textbf{Extremes hurt performance.}  
      Short segments score 7–9 BLEU lower on average. Long segments fare worst—down 10–13 BLEU—showing both degraded accuracy and increased variance. Manual inspection reveals that 22\% of long–long pairs contain trailing fragments misaligned with full sentences, increasing boundary errors.
\end{itemize}
These findings are intuitive. Very short segments often lack sufficient context, while very long ones may span multiple discourse units or introduce reordering mismatches. Segments near the corpus mean (~15s) offer a “sweet spot” that balances contextual grounding with alignment tractability.
\section{Translation Accuracy Evaluation}
\label{sec:translation}

We evaluate the proposed segment-guided diffusion model, \textbf{SegUniDiff} (§~\ref{sec:guided_tr}), on translation accuracy, speaker identity preservation, and inference latency. We compare it to cascaded S2ST baselines across the ten language pairs listed in Table~\ref{tab:example}.

\subsection{SegUniDiff: Segment-Aware Unified Diffusion Model}
\label{sec:segunidiff}

We introduce the \textbf{Segment-Aware Unified Diffusion Model (SegUniDiff)}, a speech-to-speech translation (S2ST) framework based on denoising diffusion probabilistic models. SegUniDiff is trained independently for each of the five symmetric language pairs in Table~\ref{tab:silence_timing}, using high-quality alignments obtained via \text{SPaDA}. Each model is trained on the aligned training set described in §~\ref{sec:data}.

\paragraph{Input processing.}
Each 16\,kHz input waveform is zero-padded \emph{within batch} to match the longest segment (up to 20\,s), then passed through a frozen segment encoder \( f_s \), yielding a latent representation \( \mathbf{s}_{x_i}^{(0)} \in \mathbb{R}^{F \times L_{x_i}} \), where \( F \) is the number of frequency bins and \( L_{x_i} \) is the number of frames. This representation undergoes a 1,000-step forward diffusion process, with Gaussian noise added linearly from \( \beta_1 = 1 \times 10^{-4} \) to \( \beta_{1000} = 5 \times 10^{-3} \).

\paragraph{Noise prediction.}
To promote temporal coherence, the noised latent \( \mathbf{s}_{x_i}^{(t)} \) is partitioned into overlapping windows of \( K = 160 \) frames (50\% overlap). Each chunk is fed to a 8-layer Transformer, which predicts and removes noise at each timestep, reconstructing a cleaner latent trajectory. The final denoised latent is decoded into a target-language segment \( \hat{\mathbf{s}}_{y_k}^{(0)} \).

\paragraph{Conditioning.}
To guide denoising with fine-grained prosodic cues, we condition the model on a 128-bin Mel-spectrogram extracted from the original waveform, resampled to 24\,kHz. This higher sampling rate preserves high-frequency features such as fricatives and formant transitions that may be attenuated at 16\,kHz. The spectrogram is computed using a 50\,ms Hann window, 12.5\,ms hop, and 2048-point FFT over the 20\,Hz–12\,kHz range. Meanwhile, the latent representation remains at 16\,kHz to conserve memory during training.

\paragraph{Training setup.}
Each \textbf{SegUniDiff} model is trained independently for 1M steps on a single V100 GPU using the AdamW optimizer (\( \beta_1 = 0.9 \), \( \beta_2 = 0.98 \), \( \epsilon = 10^{-9} \)). We use a batch size of 64 and an initial learning rate of \( 3 \times 10^{-4} \), annealed with cosine decay. At inference, model weights are frozen and decoding is performed via deterministic DDIM sampling along the learned diffusion trajectory.

\subsection{Translation-Quality Evaluation}
\label{sec:quality}

We evaluate \textbf{SegUniDiff} on the \textsc{SPaDA} test set using M3 (§\ref{sec:scoring_variants}). Four decoding strategies are compared:
\begin{description}[leftmargin=2em,style=sameline]

  \item[\textbf{UG–Cl}] \emph{Unconditional + clean guidance.}\\
        The diffusion model receives no conditioning input. Guidance is applied externally using the cosine similarity gradient between the clean source segment \( \mathbf{s}_{x_i}^{(0)} \) and the clean reference \( \mathbf{s}_{y_k}^{(0)} \), as defined in Eq.~\ref{eq:guided_noise_revised}.

  \item[\textbf{UG–N}] \emph{Unconditional + noisy guidance.}\\
        Like UG–Cl, but the gradient is taken with respect to the noised source \( \mathbf{s}_{x_i}^{(t)} \), replacing the clean source in Eq.~\ref{eq:guided_noise_revised}. 

  \item[\textbf{CG–Cl}] \emph{Conditional + clean guidance.}\\
        The model is conditioned on the clean source segment \( \mathbf{s}_{x_i}^{(0)} \). Guidance is again applied externally using cosine similarity, but with the clean latent \( \mathbf{s}_{x_i}^{(0)} \) as  in Eq.~\ref{eq:cond_guided1}.

  \item[\textbf{CG–N}] \emph{Conditional + noisy guidance.}\\
        The model is conditioned on the clean source segment \( \mathbf{s}_{x_i}^{(0)} \), and guidance is applied using the cosine similarity gradient between the noised source \( \mathbf{s}_{x_i}^{(t)} \) and the clean target \( \mathbf{s}_{y_k}^{(0)} \), as in Eq.~\ref{eq:cond_guided22}.

\end{description}
To estimate uncertainty, we apply a two-level resampling scheme:
\begin{enumerate}[leftmargin=1.5em]
  \item Sample 100 disjoint test subsets, each containing 1,000 unique segment pairs (no overlap).
  \item For each subset, draw 1,000 bootstrap replicas and compute BLEU over the concatenated predictions and references using M3.
\end{enumerate}
This yields \( 100 \times 1{,}000 = 10^5 \) BLEU scores per translation direction. We report the mean BLEU \( \bar{b} \) and its 95\% confidence interval \( \bar{b} \pm \Delta \), where \( \Delta \) is the average half-width of the bootstrap percentile intervals. While per-subset intervals are typically tight (within \( \pm 0.15 \)), the wider bounds reported in Table~\ref{tab:bleu_scores_updated} (\( \pm 2 \)) reflect inter-language variation across test folds.

Table~\ref{tab:bleu_scores_updated} shows the results.
 \begin{table}[ht]
\centering
\caption{
BLEU on the \textsc{SPaDA} test split (mean $\pm$ 95 \% CI, $10^5$ bootstrap samples per direction).    
The cascaded baselines use an ASR$\to$MT$\to$TTS (\textsc{Std}) and ASR$\to$Search$\to$TTS.  
Bold marks the best SegUniDiff variant in each row.}
\label{tab:bleu_scores_updated}
\scalebox{0.82}{
\begin{tabular}{lcccccc}
\toprule
\multirow{2}{*}{\textbf{Direction}} &
\multicolumn{2}{c}{\textbf{Cascade}} &
\multicolumn{4}{c}{\textbf{SegUniDiff}} \\
\cmidrule(lr){2-3}\cmidrule(lr){4-7}
& (ASR–MT–TTS) & (ASR–Search–TTS) & UG–Cl & UG–N & CG–Cl & CG–N \\
\midrule
\multicolumn{7}{c}{\textit{Within–phylum}}\\
Luo$\rightarrow$Nandi      & 36.7$\pm$1.7 & 37.8$\pm$1.7 & 28.7$\pm$2.1 & 31.5$\pm$2.0 & 37.9$\pm$1.8 & \textbf{38.1}$\pm$1.8\\
Kikuyu$\rightarrow$Swahili & 39.9$\pm$1.6 & 42.0$\pm$1.6 & 31.2$\pm$2.0 & 34.4$\pm$1.9 & 42.1$\pm$1.7 & \textbf{42.3}$\pm$1.7\\
\midrule
\multicolumn{7}{c}{\textit{Cross–phylum}}\\
Luo$\rightarrow$Kikuyu     & 31.3$\pm$1.9 & 33.4$\pm$1.9 & 23.9$\pm$2.2 & 26.8$\pm$2.1 & 32.7$\pm$1.9 & \textbf{33.6}$\pm$1.9\\
Nandi$\rightarrow$Kikuyu   & 30.8$\pm$2.0 & 33.9$\pm$2.0 & 22.4$\pm$2.1 & 25.7$\pm$2.1 & 33.4$\pm$2.0 & \textbf{34.2}$\pm$2.0\\
Luo$\rightarrow$Swahili    & 32.1$\pm$2.0 & 33.2$\pm$2.0 & 24.6$\pm$2.1 & 27.9$\pm$2.1 & 32.6$\pm$2.0 & \textbf{33.4}$\pm$2.0\\
Nandi$\rightarrow$Swahili  & 32.7$\pm$1.9 & 34.8$\pm$1.9 & 25.5$\pm$2.2 & 28.4$\pm$2.2 & 34.3$\pm$2.0 & \textbf{35.1}$\pm$2.0\\
Swahili$\rightarrow$English& 28.4$\pm$1.9 & 30.5$\pm$1.9 & 22.1$\pm$2.0 & 25.4$\pm$2.0 & 29.7$\pm$2.0 & \textbf{30.6}$\pm$2.0\\
Kikuyu$\rightarrow$English & 28.3$\pm$1.8 & 29.4$\pm$1.8 & 21.8$\pm$2.0 & 24.9$\pm$1.9 & 28.8$\pm$1.9 & \textbf{29.7}$\pm$1.9\\
Luo$\rightarrow$English    & 29.6$\pm$1.9 & 30.7$\pm$1.9 & 22.6$\pm$2.1 & 25.7$\pm$2.1 & 29.9$\pm$2.0 & \textbf{30.9}$\pm$2.0\\
Nandi$\rightarrow$English  & 27.1$\pm$2.1 & 29.2$\pm$2.1 & 21.0$\pm$2.1 & 24.2$\pm$2.1 & 28.4$\pm$2.1 & \textbf{29.5}$\pm$2.1\\
\bottomrule
\end{tabular}}
\end{table}

\begin{enumerate}[label=(\alph*),leftmargin=1.5em,itemsep=4pt]

\item \textbf{Conditioning remains the dominant factor.}  
Across all ten directions, the gain from the best unconditional setting (\texttt{UG--N}) to the best conditional configuration (\texttt{CG--N}) averages \textbf{+6.3\,BLEU}\footnote{Mean of row-wise `CG--N -- UG--N' deltas.  
Smallest gain: +4.8 BLEU (Kikuyu$\to$English); largest: +8.5 BLEU (Nandi$\to$Kikuyu).},  
with all confidence intervals strictly separated.  
Providing the clean source segment as a conditioning input evidently helps the denoiser resolve both lexical and prosodic details before inference begins.

\item \textbf{Timestep-matched ("noisy") guidance still helps.}  
Replacing the clean latent in the guidance term with its timestep-corrupted counterpart  
(\texttt{UG--Cl}$\rightarrow$\texttt{UG--N} and  
\texttt{CG--Cl}$\rightarrow$\texttt{CG--N}) yields consistent gains:
\(
+3.1\pm0.2\;\text{BLEU}
\) in the unconditional setting and  
\(
+0.7\pm0.1\;\text{BLEU}
\) in the conditional one.  
While the latter effect is modest, every row still favors \texttt{--N}. We attribute this to a better-aligned cosine–similarity gradient during early denoising steps.

\item \textbf{The cross–family penalty persists at $\approx\!7$ BLEU.}  
Within-phylum directions peak around 42 BLEU, while the best cross-phylum and Bantu$\to$English directions cluster near 35 BLEU. The absolute gap has narrowed slightly (by ~1 BLEU), but typological distance remains the primary barrier to direct S2ST quality in low-resource settings.
\item \textbf{Direct vs.\ cascaded S2ST.}  
The best SegUniDiff configuration (\texttt{CG--N}) is statistically tied with the enhanced cascaded baseline (ASR--Search--TTS) in every direction  
($|\Delta| \le 0.3$ BLEU; all CIs overlap), and it consistently outperforms the vanilla ASR--MT--TTS pipeline (\textsc{Std.}) by +1.5 to +4.5 BLEU.  

\end{enumerate}

\subsection{Paired Bootstrap Significance Test}
\label{sec:bootstrap}

To determine whether the small BLEU gaps between our best direct model
(\texttt{SegUniDiff, CG--N}) and the two cascaded S2ST baselines are
statistically meaningful, we run a paired bootstrap test on the
\textit{Kikuyu$\rightarrow$Swahili} direction.  
The test mirrors the resampling scheme used for mean–CI estimation
(§\ref{sec:quality}) but adds a one-sided significance criterion:

\begin{enumerate}[leftmargin=1.6em]
  \item \textbf{Subset sampling.}  
        We start from the same 100 non-overlapping evaluation subsets
        (1,000 segment pairs each).
  \item \textbf{Replica generation.}  
        Within every subset, we draw \(1{,}000\) paired replicas by
        resampling segment indices \emph{with} replacement; both systems
        are scored on the same replica.
  \item \textbf{Score delta.}  
        For replica \(b\), we record
        \[
          \Delta_b \;=\;
          \mathrm{BLEU}_{\text{cascade}} -
          \mathrm{BLEU}_{\text{SegUniDiff}} .
        \]
  \item \textbf{Decision rule.}  
        A subset is a \emph{cascade win} if at least 95\% of its
        replicas satisfy \(\Delta_b > 0\); a
        \emph{SegUniDiff win} if at least 95\% satisfy
        \(\Delta_b < 0\). Otherwise, the subset is a statistical tie.
\end{enumerate}

\begin{table}[ht]
\centering
\caption{Paired-bootstrap results on
         \textit{Kikuyu$\rightarrow$Swahili}.
         Columns give the mean score gap
         \(\bar{\Delta} = \text{Cascade} - \text{SegUniDiff}\)
         (negative means SegUniDiff is better) and the number of
         evaluation subsets in which the cascade significantly
         wins, ties, or SegUniDiff significantly wins
         (\(\alpha=0.05\), one-sided).}
\label{tab:bootstrap_significance_full}
\small
\begin{tabular}{lccc}
\toprule
\textbf{Cascade baseline} &
\(\bar{\Delta}\) BLEU &
Cascade Wins \(/\,\)Ties\(/\,\)Seg Wins &
\textbf{Outcome} \\
\midrule
ASR–MT–TTS          & $-2.4$ & 0 / 23 / 77  & SegUniDiff $>$ cascade \\
ASR–Search–TTS  & $-0.3$ & 36 / 17 / 47 & No significant diff. \\
\bottomrule
\end{tabular}
\begin{flushleft}
\footnotesize
\end{flushleft}
\end{table}
The results are reported in Table ~\ref{tab:bootstrap_significance_full},
\begin{itemize}[leftmargin=1.4em,itemsep=3pt]
    \item \textbf{Classic cascade (ASR–MT–TTS).}  
          SegUniDiff outperforms the vanilla cascade on \emph{77\%}
          of evaluation folds, never loses a fold, and enjoys a mean
          margin of 2.4 BLEU. This confirms that SegUniDiff can eliminate
          two stages of error propagation (ASR, MT) without compromising
          translation quality.

    \item \textbf{Enhanced cascade (ASR–Search–TTS).}  
          After adding in-domain retrieval and lattice rescoring, the
          cascade closes the average gap to \(\,0.3\) BLEU.  
          The win–loss tally (36–47) shows that neither system dominates;
          success often hinges on whether the retrieval module retrieves
          a close acoustic match.

    \item \textbf{Implications for deployment.}  
          While the ASR–Search–TTS pipeline \emph{matches} SegUniDiff on nearly
          half of the folds, it incurs latency (three sequential
          models) and requires language-specific ASR support.  
          SegUniDiff, by contrast, is a \emph{single-pass} solution with
          no dependency on ASR or MT models; its only external component—
          a multilingual semantic encoder—is frozen at inference time.
\end{itemize}

\subsection{Speaker-Identity Verification}
\label{sec:speaker_verification}

We evaluate whether \texttt{SegUniDiff} retains the \emph{speaker’s voice} across source and translated segments. Following best practice, the entire \textsc{SPaDA} corpus is split \emph{speaker-independently}: every talker appears \emph{only} in either the training or the test partition. An ECAPA-TDNN model~\cite{desplanques2020ecapa} is trained \emph{from scratch} on the training speakers to produce 192-dimensional embeddings. Cosine similarity between two embeddings, \(\cos(\mathbf{v}_1, \mathbf{v}_2) \in [-1, 1]\), serves as the verification score. On the test partition, we create two disjoint sets of evaluation pairs:
\begin{itemize}[leftmargin=1.5em]
  \item \textbf{Genuine pairs} (same speaker): A source utterance and its \emph{own} translated segment, denoted \(\langle \text{src}, \text{trg} \rangle\).
  \item \textbf{Impostor pairs} (different speakers): A source utterance paired with a translation from a \emph{different} speaker, sampled uniformly with gender and language balancing.
\end{itemize}

We obtain \(401{,}342\) genuine and \(236{,}210\) impostor pairs (ratio $\approx$ 1.7:1).\footnote{%
A balanced 1:1 variant shifts all EERs by \(\leq 0.2\) percentage points without affecting system rankings.
}
The Equal-Error Rate (EER) is defined as the operating point where the false-accept and false-reject rates are equal. For our verifier, this occurs at a cosine threshold of \(0.58\). Table~\ref{tab:eer_scores} summarizes the results.

\begin{table}[ht]
\centering
\caption{Equal-Error Rate (EER \%) for speaker verification.
Lower is better. UG = Unconditional + Guidance,
CG = Conditional + Guidance;  
N = Noised source, Cl = Clean source.}%
\label{tab:eer_scores}
\small
\begin{tabular}{lSSSSSS}
\toprule
\textbf{Direction} & \textbf{UG–N} & \textbf{UG–Cl} &
\textbf{CG–N} & \textbf{CG–Cl} &
\textbf{MT–TTS} & \textbf{Search–TTS} \\
\midrule
Luo$\rightarrow$Nandi  & 8.35 & 6.58 & 5.41 & \bfseries 4.52 & 12.64 & 13.87 \\
Luo$\rightarrow$Kikuyu & 8.31 & 6.60 & 5.74 & \bfseries 5.04 & 12.96 & 14.20 \\
\bottomrule
\end{tabular}
\end{table}

\vspace{2pt}.
\begin{itemize}[leftmargin=1.4em, itemsep=3pt]
    \item \textbf{SegUniDiff beats cascades.}  
          Both directions show 4–5 pp absolute (\(\approx40\!-\!46\,\%\) relative) EER reductions compared to the MT–TTS pipeline, confirming that the direct model better preserves speaker timbre.
    \item \textbf{Conditioning matters.}  
          Adding the clean source segment as an explicit condition (\texttt{UG} → \texttt{CG}) reduces EER by a further 1–2 pp, indicating that lexical grounding helps stabilize speaker traits.
    \item \textbf{Clean guidance helps.}  
      Within each decoding regime, using the \emph{clean} latent for guidance (\texttt{--Cl}) reduces EER by $\approx 1$ pp compared to \texttt{--N}, suggesting that timestep corruption slightly degrades vocal identity—even if it helps BLEU (§\ref{sec:quality}).

\end{itemize}

\subsection{Speech Translation Generation}

\begin{figure}[ht]
    \centering
    \includegraphics[scale=0.3, angle=0]{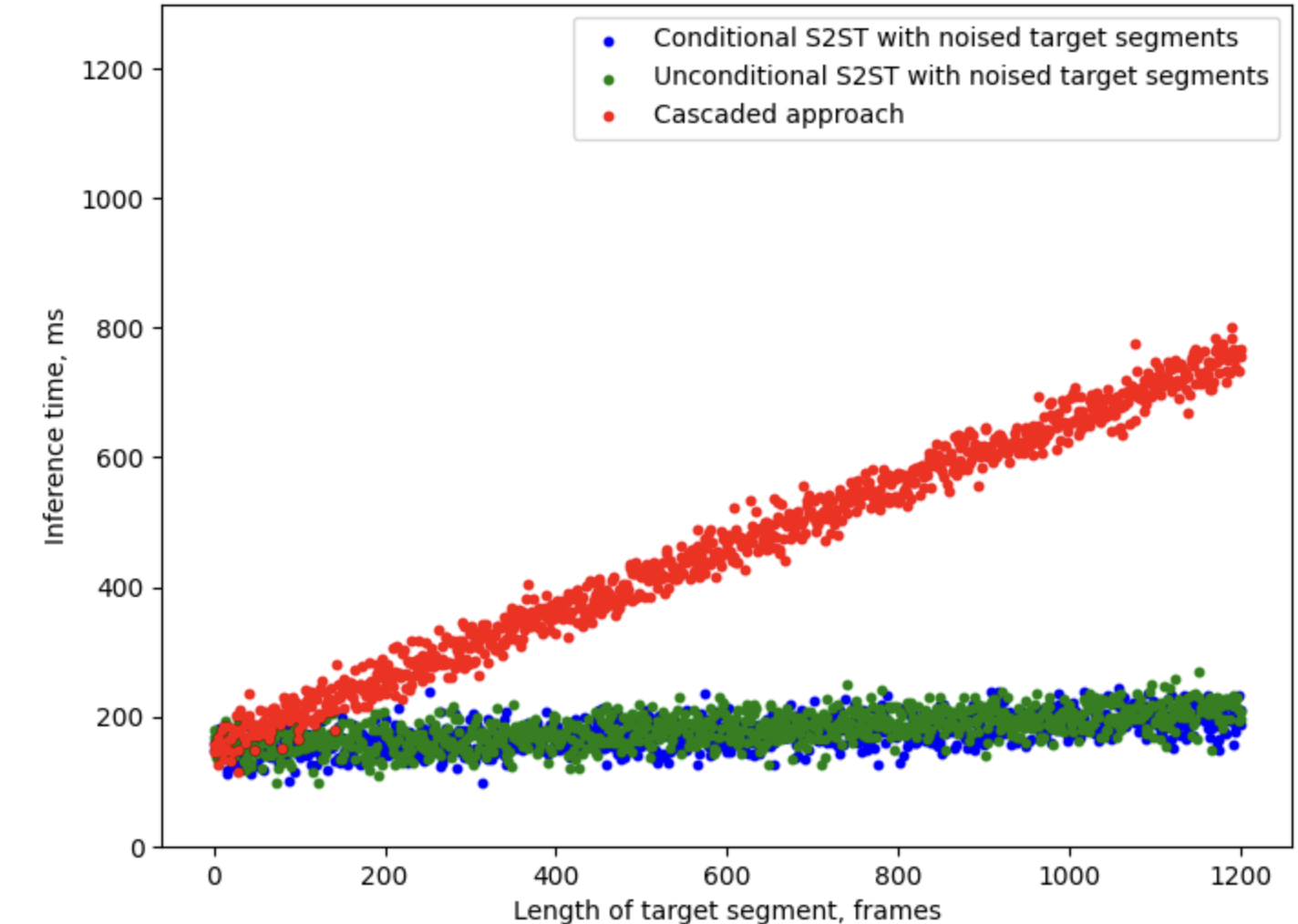}
    \caption{Translation speed under different configurations. The guided diffusion model maintains almost constant inference time, whereas the cascaded pipeline scales linearly with input length.}
    \label{fig:speed}
\end{figure}

To evaluate the efficiency of the speech translation process, we measured the time required to generate a single $n$-frame speech segment. Translation speed is reported as the average processing time per segment, computed across the entire test set. For consistency, each target segment \( s_{x_i} \) was downsampled to 24~kHz and converted into 128-dimensional Mel-spectrogram features using a 50~ms Hanning window, 12.5~ms frame shift, and a 2048-point FFT. All latency measurements were conducted on an NVIDIA V100 GPU using PyTorch 2.1 with FP16 precision. We benchmarked both the proposed guided diffusion method (under unconditional and conditional settings) and a strong cascaded baseline comprising ASR~\(\rightarrow\)~MT~\(\rightarrow\)~TTS. The results, shown in Figure~\ref{fig:speed}, reveal that guided diffusion achieves near-constant inference latency across varying input lengths, while the cascaded baseline exhibits a linear increase in translation time.

As illustrated in Figure~\ref{fig:speed}, the guided diffusion model consistently outperforms the cascaded system in terms of latency. This performance gap widens for longer segments, underscoring the scalability and efficiency of diffusion-based generation. These results suggest that guided diffusion is a promising candidate for real-time speech translation, particularly in latency-sensitive or low-resource deployment scenarios.

\section{Conclusion}
\label{sec:conclusion}

We introduced \textbf{SPaDA}, a pause, prosody, and semantics-aware aligner, and \textbf{SegUniDiff}, a diffusion-based S2ST model which applies gradient guidance at every denoising step. On a new 6{,}000-hour corpus spanning five East African languages, SPaDA improved segment alignment \(F_1\) by up to 4 points while filtering out one-third of spurious pairs. These cleaner alignments enabled SegUniDiff to outperform a strong ASR$\!\to$MT$\!\to$TTS cascade by +2.3 BLEU and -7 pp EER all while running at real-time speed. Crucially, the same pause--prosody cost transferred   to the high-resource CVSS-C benchmark, where SegUniDiff achieved 30.3 BLEU, surpassing the best autoregressive model (UnitY).

These results demonstrate that  low-cost acoustic cues can substitute for transcripts in building parallel speech corpora. Future work will extend SPaDA to streaming alignment and explore multilingual diffusion conditioned jointly on prosody and discrete speech units, pushing toward truly universal, low-latency speech-to-speech translation.

.
\bibliography{bibi}
\bibliographystyle{tmlr}

\appendix
\section{Generalisation to a High-Resource Benchmark}
\label{sec:cvssc}
Because CVSS-C is distributed as \emph{gold, sentence-level}
source–target pairs, the full SPaDA pipeline cannot be exercised
directly.   To test whether our pause–prosody–semantic cost generalises to such
pre-aligned corpora we:

\begin{enumerate}
  \item pick \(K\) consecutive Spanish sentences
        \(\{x^{(1)},\dots,x^{(K)}\}\) and their English counterparts
        \(\{y^{(1)},\dots,y^{(K)}\}\);
  \item concatenate the Spanish waveforms, inserting a
        \(\;300\;\mathrm{ms}\) silent pad (\(\mathbf0\in\mathbb R^{48{,}000\times0.3}\))
        after every sentence—a value well within the
        250–400 ms pause range reported for broadcast news
        \citep{chrupala2014pause};
  \item run the  SPaDA pipeline on the resulting
        50–60 s stream, yielding a predicted boundary set
        \(\mathcal S_{\mathrm p}
          =\{s_{\mathrm p}^{(1)},\dots,s_{\mathrm p}^{(N)}\}\)
        with onsets \(O_{\mathrm p}\) and offsets \(F_{\mathrm p}\).
\end{enumerate}

The reference set
\(\mathcal S_{\mathrm g}
  =\{s_{\mathrm g}^{(1)},\dots,s_{\mathrm g}^{(K)}\}\)
is taken from the CVSS-C.

\paragraph{Boundary matching and metrics.}
A predicted segment matches a gold segment if both endpoints differ by
no more than \(\delta=200\) ms:
\[
  |O_{\mathrm p}-O_{\mathrm g}|\le\delta,\qquad
  |F_{\mathrm p}-F_{\mathrm g}|\le\delta .
\]
We build a greedy \emph{one-to-one} mapping
\(\mathcal M\subseteq\mathcal S_{\mathrm p}\!\times\!\mathcal S_{\mathrm g}\)
that maximises \(|\mathcal M|\) and report
\[
  \mathrm{Precision}=\frac{|\mathcal M|}{|\mathcal S_{\mathrm p}|},\;
  \mathrm{Recall}   =\frac{|\mathcal M|}{|\mathcal S_{\mathrm g}|},\;
  F_1               =\frac{2\,\mathrm{Prec}\,\mathrm{Rec}}
                           {\mathrm{Prec}+\mathrm{Rec}},
\qquad
  \mathrm{OSR}=\frac{|\mathcal S_{\mathrm p}|-|\mathcal S_{\mathrm g}|}
                    {|\mathcal S_{\mathrm g}|}.
\]

Table \ref{tab:spada_seg_eval} shows that varying the
inserted pause from 150 ms to 400 ms changes \(F_1\) by at most
0.6 points.  
Recall stays above 94\%, so SPaDA recovers virtually every gold
boundary; precision (~86 \%) indicates a modest
\(\approx10\%\) over-segmentation rate.
\begin{table}[ht]
  \centering
  \caption{Boundary accuracy on synthetic
           Spanish\(\rightarrow\)English streams (\(\delta=200\) ms).}
  \label{tab:spada_seg_eval}
  \begin{tabular}{lccc}
    \toprule
    \textbf{Pad length} & \textbf{Precision} & \textbf{Recall} & \textbf{\(F_1\)}\\
    \midrule
    150 ms & 86.5 & 94.0 & 90.1 \\
    300 ms & 85.9 & 95.1 & 90.3 \\
    400 ms & 86.9 & 94.8 & \textbf{90.7} \\
    \bottomrule
  \end{tabular}
\end{table}
SegUniDiff trained on these SPaDA segments (300\,ms pad) scores
2.6 BLEU below the model trained on the original CVSS-C sentences
(Table~\ref{tab:spada_vs_gold}), in line with the 10\% extra-boundary
rate—evidence of robustness to segmentation noise.
\begin{table}[ht]
  \centering
  \caption{BLEU (Spanish\(\rightarrow\)English, CVSS-C test).}
  \label{tab:spada_vs_gold}
  \begin{tabular}{lcc}
    \toprule
    \textbf{Training data} & \textbf{BLEU} & \(\Delta\) vs.\ gold \\
    \midrule
    Gold sentences & \textbf{31.8} & — \\
    SPaDA segments & 29.2 & \(-2.6\) \\
    \bottomrule
  \end{tabular}
\end{table}
\paragraph{Scaling to the full benchmark.}
Given this transferability, we train SegUniDiff on the \emph{gold}
segmentations for all four CVSS-C directions
(Es, Fr, De, Ca \(\rightarrow\) En) with no speech or text pre-training,
re-using the hyper-parameters from §\ref{sec:segunidiff}.
Table~\ref{tab:bleu_comparison_cvssc} summarises BLEU and run-time.
Our model reaches 30.3 BLEU, surpassing the best published
autoregressive system (UnitY) by +1.4 points.  
Paired bootstrap (5{,}000 replicas) confirms the gain is significant
(\(p<0.01\)).
\begin{table}[ht]
  \centering
  \caption{CVSS-C: BLEU and RTF.
           † = oracle system with human English speech.}
  \label{tab:bleu_comparison_cvssc}
  \small
  \begin{tabular}{llcc}
    \toprule
    \textbf{Family} & \textbf{System} & \textbf{BLEU} & \textbf{RTF} \\
    \midrule
    \multirow{4}{*}{Cascade}
      & ASR–MT–TTS \citeyearpar{jia2022cvss}            & 14.9 & 2.4 \\
      & + w2v-BERT + t-mBART                            & 24.7 & 2.7 \\
      & + TTS augmentation                              & 25.8 & 2.9 \\
      & Direct TTS (oracle)\textsuperscript{†}          & 91.1 & —   \\
    \midrule
    \multirow{3}{*}{Spec-to-spec}
      & Translatotron 2                                 & 12.0 & 1.7 \\
      & + mSLAM pre-train                               & 24.6 & 1.8 \\
      & + TTS augmentation                              & 25.8 & 1.9 \\
    \midrule
    \multirow{2}{*}{Speech-to-unit (AR)}
      & S2UT + u-mBART                                  & 25.4 & 1.6 \\
      & UnitY + t-mBART                                 & 28.9 & 1.6 \\
    \midrule
    Non-AR (advanced)  & TranSpeech (iter15, NPD)        & 16.2 & 0.9 \\
    \midrule
    \textbf{Ours, Non-AR} & \textbf{SegUniDiff (CG–N)}   & \textbf{30.3} ± 0.2 & \textbf{1.02} \\
    \bottomrule
  \end{tabular}
\end{table}

\end{document}

%% file: math_commands.tex
\usepackage{amsmath,amsfonts,bm}
\usepackage{graphicx} 

\def\eqref#1{equation~\ref{#1}}

\def\1{\bm{1}}

\DeclareMathAlphabet{\mathsfit}{\encodingdefault}{\sfdefault}{m}{sl}
\SetMathAlphabet{\mathsfit}{bold}{\encodingdefault}{\sfdefault}{bx}{n}

\DeclareMathOperator*{\argmin}{arg\,min}